 \documentclass[twocolumn]{aastex63}

\shorttitle{Maximum Magnitude versus Rate of Decline}
\shortauthors{Hachisu et al.}

\begin{document}

\title{A Theory for the Maximum Magnitude versus Rate of Decline
(MMRD) Relation of Classical Novae}


\author{Izumi Hachisu}
\affil{Department of Earth Science and Astronomy, 
College of Arts and Sciences, The University of Tokyo,
3-8-1 Komaba, Meguro-ku, Tokyo 153-8902, Japan} 
%%%\email{hachisu@ea.c.u-tokyo.ac.jp}
\email{izumi.hachisu@outlook.jp}

%\and

\author{Hideyuki Saio}
\affil{Astronomical Institute, Graduate School of Science,
    Tohoku University, Sendai 980-8578, Japan}
% \email{saio@astr.tohoku.ac.jp}

%\and

\author{Mariko Kato}
\affil{Department of Astronomy, Keio University, 
Hiyoshi, Kouhoku-ku, Yokohama 223-8521, Japan} 
%%\email{mariko@educ.cc.keio.ac.jp}

%\and

\author{Martin Henze}
\affil{Department of Astronomy, San Diego State University, San Diego, 
CA 92182, USA}

\author{Allen W. Shafter}
\affil{Department of Astronomy, San Diego State University, San Diego, 
CA 92182, USA}


%\altaffiltext{1}{Visiting Astronomer, Cerro Tololo Inter-American Observatory.
%CTIO is operated by AURA, Inc.\ under contract to the National Science
%\altaffiltext{2}{Society of Fellows, Harvard University.}
%\altaffiltext{3}{present address: Center for Astrophysics,
%    60 Garden Street, Cambridge, MA 02138}
%\altaffiltext{4}{Visiting Programmer, Space Telescope Science Institute}
%\altaffiltext{5}{Patron, Alonso's Bar and Grill}


\begin{abstract}
We propose a theory for the MMRD relation of novae, using free-free emission
model light curves built on the optically thick wind theory.
We calculated $(t_3,M_{V,\rm max})$ for various sets of $(\dot M_{\rm acc},
M_{\rm WD})$, where $M_{V,\rm max}$ is the peak absolute $V$ magnitude,
$t_3$ is the 3-mag decay time from the peak, and $\dot M_{\rm acc}$ is
the mass accretion rate on to the white dwarf (WD) of mass $M_{\rm WD}$.
The model light curves are uniquely characterized 
by $x\equiv M_{\rm env}/M_{\rm sc}$, where $M_{\rm env}$ is the
hydrogen-rich envelope mass and $M_{\rm sc}$ is the scaling mass
at which the wind has a certain wind mass-loss rate.
For a given ignition mass $M_{\rm ig}$, we can specify the first point 
$x_0= M_{\rm ig}/M_{\rm sc}$ on the model light curve, and calculate 
the corresponding peak brightness and $t_3$ time from this first point.  
Our $(t_3, M_{V,\rm max})$ points cover well the distribution 
of existing novae.  The lower the mass accretion rate, the brighter 
the peak.  The maximum brightness is limited to 
$M_{V,\rm max}\gtrsim -10.4$ by the lowest mass-accretion rate of 
$\dot M_{\rm acc}\gtrsim1\times10^{-11}~M_\sun$~yr$^{-1}$.   
A significant part of the observational MMRD trend corresponds to 
the $\dot M_{\rm acc}\sim5\times10^{-9}~M_\sun$~yr$^{-1}$ line
with different WD masses.
A scatter from the trend line indicates a variation in their
mass-accretion rates.
Thus, the global trend of an MMRD relation does exist, but 
its scatter is too large for it to be a precision distance indicator
of individual novae.
%We tabulate $(t_3, M_{V,\rm max})$ for many sets of
%$(\dot M_{\rm acc},M_{\rm WD})$. 
\end{abstract}




\section{Introduction}
\label{introduction}
A typical classical nova shows a rapid rise of optical brightness
until its peak followed by a slow decline.  There is a statistical trend
that a faster decline nova shows a brighter optical peak. 
The scatter around the main trend, however, is not small.  
It has long been debated whether or not a meaningful relation actually
exists \citep[e.g.,][]{mcl45, sch57, coh85, del95,
dow00, kas11, shafter11, cao12,
shafter13, shara17, schaefer18, ozd18, sel19, del20}.  
Such a relation is called the maximum magnitude versus rate of decline 
(MMRD) relation.  If an MMRD relation exists and is a simple monotonic
relation, it can be used to obtain the absolute peak brightness of a nova 
from the rate of decline, and thus, it would be a useful tool to obtain 
the distance to a nova.  Although there are early attempts to theoretically
explain the MMRD relation \citep[e.g.,][]{liv92},
we need a convincing theoretical background
to understand the main trend and large scatter of the existing MMRD 
distribution of novae.

A nova is a thermonuclear runaway event on a mass-accreting white dwarf (WD).
Hydrogen ignites and releases nuclear energy.  The hydrogen-rich envelope 
expands to a giant size.  The subsequent nova evolution was
theoretically followed by \citet{kat94h} based on the assumption of
spherical symmetry.  Strong optically-thick winds are accelerated
deep inside the photosphere.  The wind stops after a significant
part of the hydrogen-rich envelope is ejected by the wind.
The timescale of a nova in the early phase is determined by the wind
mass-loss rate and the amount of the hydrogen-rich envelope mass.  
%
%The nova early brightness is determined by flux of free-free emission
%because free-free emission dominates the spectra of novae 
%\citep[e.g.,][]{enn77, gal76}.  
%The free-free emission brightness depends mainly on the wind mass-loss rate
%and therefore the nova brightness begins to decline 
%after the maximum wind mass-loss
%rate is attained \citep[e.g.,][]{hac06kb, hac17k}.

\citet{kat94h} calculated optically thick winds 
in the decay phase of novae and obtained the photospheric 
radius $R_{\rm ph}$, temperature $T_{\rm ph}$, luminosity
$L_{\rm ph}$, velocity $v_{\rm ph}$, and wind mass-loss rate 
$\dot M_{\rm wind}$ against the decreasing envelope mass $M_{\rm env}$.
Observationally, early spectra of novae are dominated by free-free emission
\citep[e.g.,][]{enn77, gal76}.  Therefore, \citet{hac06kb} calculated 
free-free emission model light curves with $F_\nu \propto 
\dot M_{\rm wind}^2 / (v_{\rm ph}^2 R_{\rm ph})$,
where $F_\nu$ is the flux at the frequency $\nu$.
These model light curves well reproduce many nova light curves 
\citep[e.g.,][]{hac06kb, hac10k, hac16k, hac18kb, hac19ka, hac19kb}.

The theoretical free-free emission light curves show a homologous shape
independent of the WD mass and chemical composition.  
\citet{hac06kb} called this property of
nova model light curves ``the universal decline law.''
These properties, i.e., homologous and frequency independent shapes of
light curves, indicate that the model light curves are expressed
by a unique function of a parameter.  
We find that this parameter is the ratio of the envelope mass 
and the scaling envelope mass having a certain wind mass-loss rate, that is, 
$x\equiv M_{\rm env}/M_{\rm sc}$ as will be explained in Section 
\ref{timescaling_law_free-free_emission}.  

\citet{hac10k} found that two different model light curves, 
e.g., corresponding to the two different WD masses, 
can overlap each other if the timescale of one of them is squeezed by 
a factor of $f_{\rm s}$, i.e., $t/f_{\rm s}$.  
The normalization factor is $f_{\rm s}< 1$ for a faster nova 
(corresponding to a more massive WD), and $f_{\rm s}> 1$ for a slower nova 
(a less massive WD).  Then the absolute $V$ brightnesses is normalized to be
$M_V-2.5\log f_{\rm s}$.  Thus, the two different light curves 
overlap each other in the $(t/f_{\rm s})$--$(M_V-2.5\log f_{\rm s})$ 
plane \citep[see, e.g., Figures 48 and 49 of][]{hac18kb}.  
\citet{hac19kb} reformulated this property:
if the $V$ light curve of a template nova (time $t$) overlaps with 
that of a target nova (time $t'=t/f_{\rm s}$), we have the relation
\begin{eqnarray}
\left( M_V[t] \right)_{\rm template} 
&=& \left( M'_V[t'] \right)_{\rm target} \cr
&=& \left( M_V[t/f_{\rm s}]-2.5\log f_{\rm s} \right)_{\rm target},
\label{time-stretching_general}
\end{eqnarray}
where $M_V[t]$ is the original absolute $V$ brightness and $M'_V[t']$ 
is the time-normalized brightness after time-normalization
of $t'=t/f_{\rm s}$.
This property was calibrated on many novae \citep[e.g.,][]{hac16k,
hac18kb, hac19ka}.

\citet{hac10k} also presented a theoretical explanation of the MMRD relation 
based on the universal decline law and Equation 
(\ref{time-stretching_general}).  Their interpretation on the main trend
of the MMRD distribution is that V1668 Cyg is a typical classical nova,
and that the novae having the same time-normalized light curves 
(i.e., the same normalized peak brightnesses) as that of V1668~Cyg but
the different WD masses form the main trend line in the 
$(\log t_3)$--$M_{V, \rm max}$ diagram, i.e.,
$M_{V, \rm max}= 2.5 \log t_3 -11.94$ \citep[Equation (25) in ][]{hac16k}.
Here, $t_3$ is the 3-mag decay
time from the $V$ peak and $M_{V, \rm max}$ is the absolute
$V$ peak magnitude. 
This main trend line is located in the middle of the observational
MMRD distribution of novae in the $(\log t_3)$--$M_{V, \rm max}$ diagram
and its peak becomes brighter along Equation 
(\ref{time-stretching_general}) with the decreasing $f_{\rm s}$.
On this main trend line, a faster decline nova with a shorter $t_3$ time 
(smaller $f_{\rm s}$) corresponds to a more massive WD while a slower 
decline nova with a longer $t_3$ time (larger $f_{\rm s}$) 
does to a less massive WD. 

In the present work, we clarify the physics of MMRD points.
Then, we explain the main trend of MMRD relation as a typical 
$\dot M_{\rm acc}$ with the different $M_{\rm WD}$'s.
The scatter from the main trend line is explained by the difference
of $\dot M_{\rm acc}$ from a typical $\dot M_{\rm acc}$. 

Our paper is organized as follows.  First we propose
several timescaling laws
and clarify the physics of MMRD relation in Section 
\ref{timescaling_law_free-free_emission}.
Then, we approximate these timescaling laws with analytic
expressions in Section \ref{approximate_relation},
which simplify the calculations of $(t_3, M_{V, \rm max})$.
In Section \ref{theoretical_mmrd_relation}, we explain our theoretical 
$(t_3, M_{V, \rm max})$ relation on the base of 
$(\dot M_{\rm acc}, M_{\rm WD})$, the main characteristic properties of
cataclysmic binaries.
Discussion and our conclusions are given in Sections 
\ref{discussion} and \ref{conclusions}, respectively.  
We tabulate our numerical results in
Appendix \ref{accreted_envelope_ignition}.

\begin{figure*}
\epsscale{1.1}
\plotone{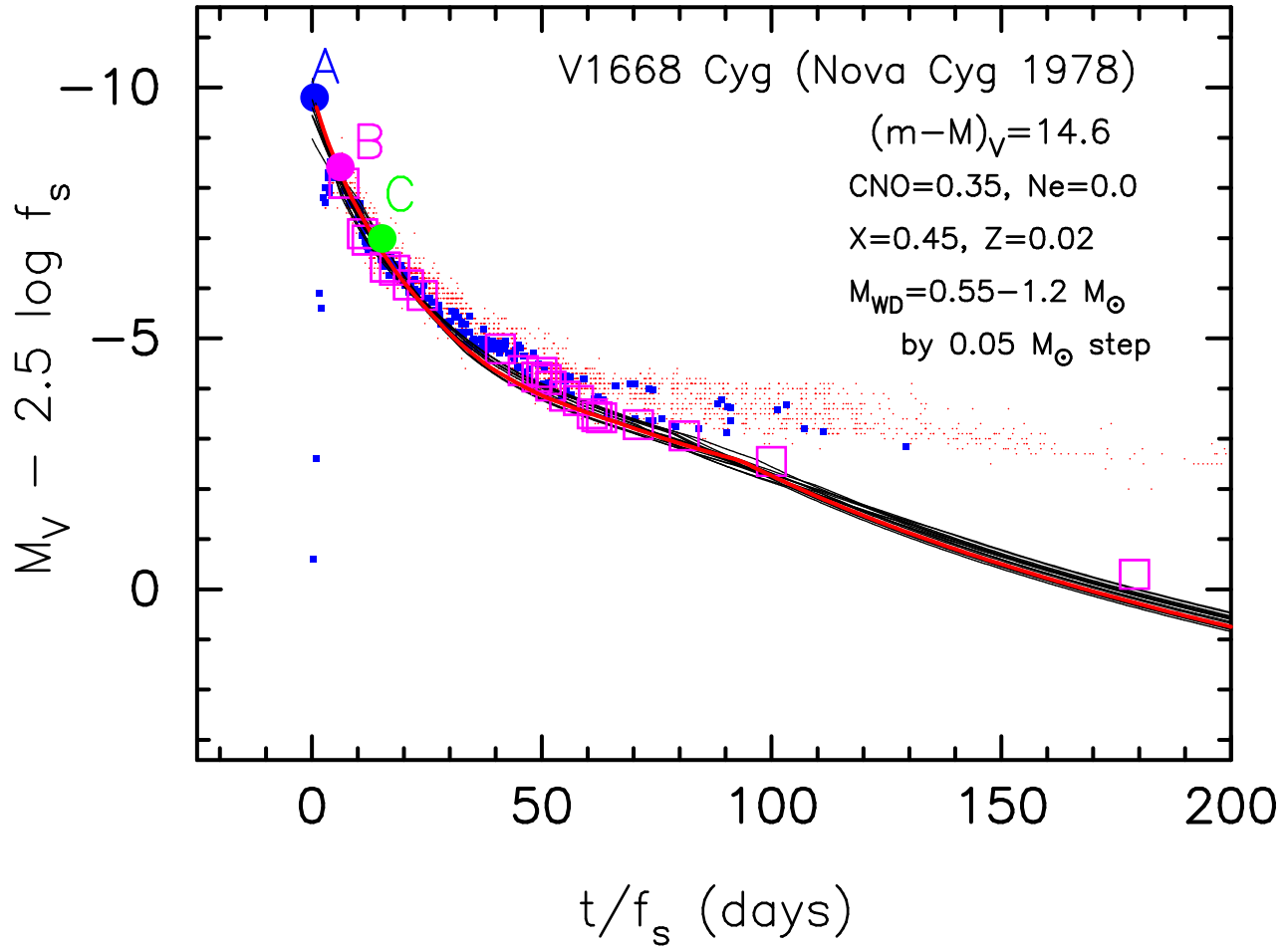}
%%\plottwo{f4a.eps}{f4b.eps}
%%%%\plotone{all_mass_v1668_cyg_x45z02c15o20_calib_universal_linear_no3.epsi}
%\plotfiddle{evolution1.ps}{5.0cm}{270}{0.4}{0.4}{-170}{220}
\caption{
Comparison of our model $V$ light curves with the V1668~Cyg light curve.
A $0.98 ~M_\sun$ WD (CO3) model (solid red line) reasonably reproduces
the V1668~Cyg optical $y$ (unfilled magenta squares) 
and $V$ (filled blue squares) light curves.  
We add the visual magnitudes (red dots) of V1668~Cyg.
The distance modulus in $V$ band of $\mu_V\equiv (m-M)_V= 14.6$ is
taken from \citet{hac19ka}.  
The data are all the same as Figure 46 of \citet{hac18kb}.
Free-free emission model $V$ light curves
for 0.55, 0.60, 0.65, 0.70, 0.75, 0.80,
0.85, 0.90, 0.95, 1.0, 1.05, 1.1, 1.15, and $1.20~M_\sun$ WDs
(solid black lines) are plotted 
in the $(t/f_{\rm s})$-$(M_V-2.5\log f_{\rm s})$ plane.
The time-normalization factor $f_{\rm s}$ of each model, tabulated
in Table 3 of \citet{hac16k}, is measured against that
of the V1668~Cyg light curves.
We place three points, A, B, and C, on the model $V$ light curve
($0.98~M_\sun$),
corresponding to three different initial envelope masses,
$M_{\rm env,0} = 1.8 \times 10^{-5}~M_\sun$, $1.4 \times 10^{-5}~M_\sun$,
and $0.94 \times 10^{-5}~M_\sun$, respectively.
Point B is the optical peak of V1668~Cyg, 
$m_{V, \rm max}=6.2~(M_{V, \rm max}=6.2 - 14.6=-8.4)$.
\label{all_mass_v1668_cyg_x45z02c15o20_calib_universal_linear_no3}}
\end{figure*}

\section{Timescaling Law of Free-Free Emission Model Light Curves}
\label{timescaling_law_free-free_emission}

\begin{figure*}
\epsscale{1.1}
%%\rotate
%%\plotone{f317.eps}
\plotone{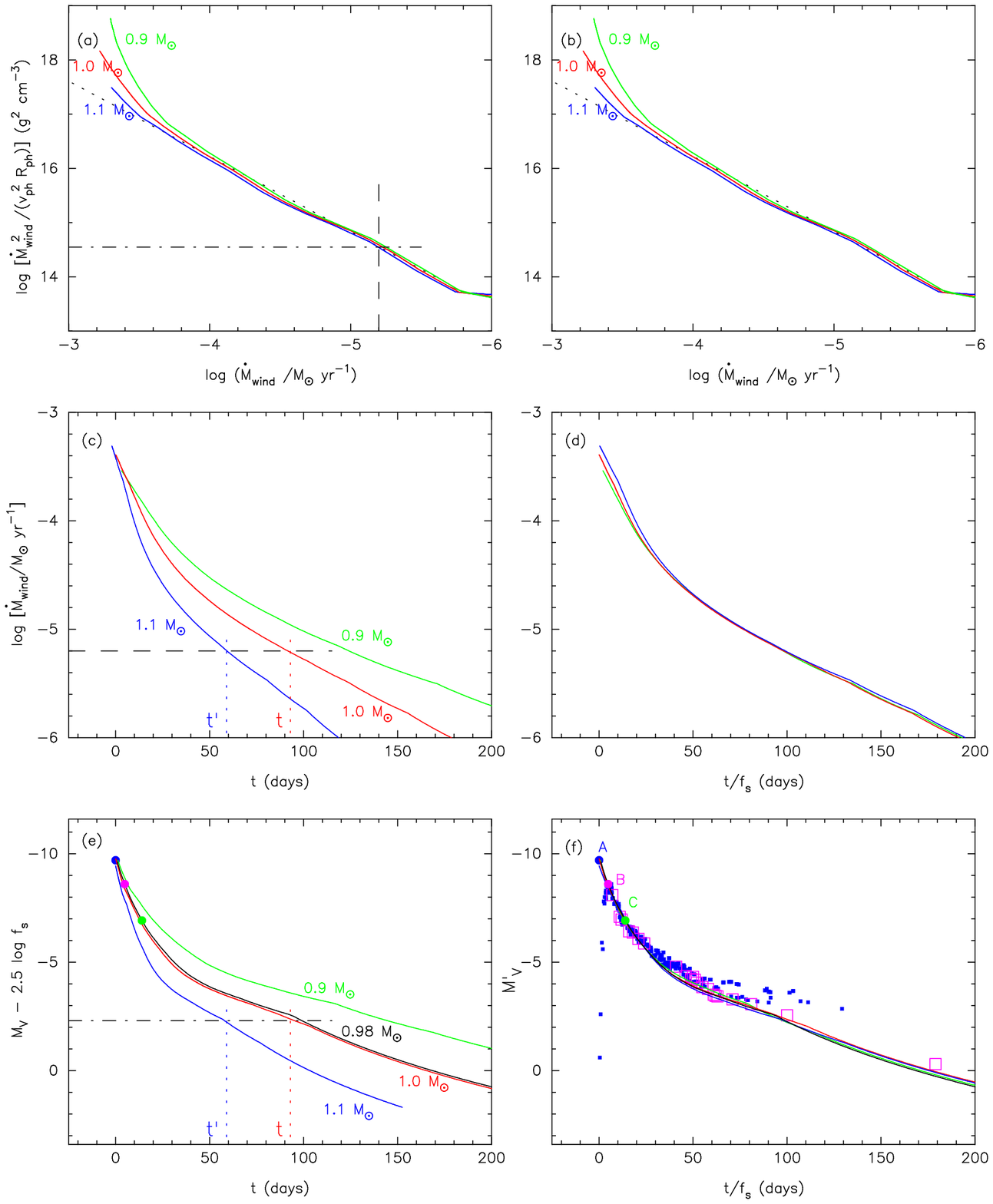}
%\plotone{dmdt_ff-flux_env_mag_relation_x45z02c15o20.epsi}
%\plotone{dmdt_ff_flux_t_only_6fig_x45z02c15o20_no2.epsi}
%\plotfiddle{evolution1.ps}{5.0cm}{270}{0.4}{0.4}{-170}{220}
\caption{
(a)(b) 
The free-free flux parameter of 
$\dot M_{\rm wind}^2 / (v_{\rm ph}^2 R_{\rm ph})$
versus wind mass-loss rate $\dot M_{\rm wind}$ 
for $1.1~M_\sun$ (blue), $1.0~M_\sun$ (red), and $0.9~M_\sun$ (green)
WD (CO3) models, which are the same models as in Figure
\ref{all_mass_v1668_cyg_x45z02c15o20_calib_universal_linear_no3}.
The three (blue, red, and green) lines almost overlap each other. 
The dotted lines indicate the global trend of these three lines.
(c)(d) 
The wind mass-loss rates are plotted against the real time $t$ 
and the normalized time $t/f_{\rm s}$, respectively, 
for the same three WD mass models.
(e)(f) 
Free-free emission model $V$ light curves for the $0.90 ~M_\sun$ (green), 
$0.98 ~M_\sun$ (black), $1.0 ~M_\sun$ (red), and $1.1~M_\sun$ (blue) WDs 
are plotted against the real time $t$ and the normalized time $t/f_{\rm s}$,
respectively.  In panel (e), the right edge of $1.1~M_\sun$ light
curve (blue line) corresponds to the end of wind phase.
Note that the ordinate in panel (f) is
$M'_V[t']$ in Equation (\ref{time-stretching_general}),
corresponding to the abscissa of $t'=t/f_{\rm s}$ while the ordinate
in panel (e) is $M_V[t]-2.5\log f_{\rm s}$ because the abscissa is
$t$.  The symbols in panels (e) and (f) are the same as those in Figure
\ref{all_mass_v1668_cyg_x45z02c15o20_calib_universal_linear_no3}. 
See the text for more details. 
\label{dmdt_ff-flux_env_mag_relation_x45z02c15o20}}
\end{figure*}

\begin{figure*}
\epsscale{1.15}
%%\rotate
\plotone{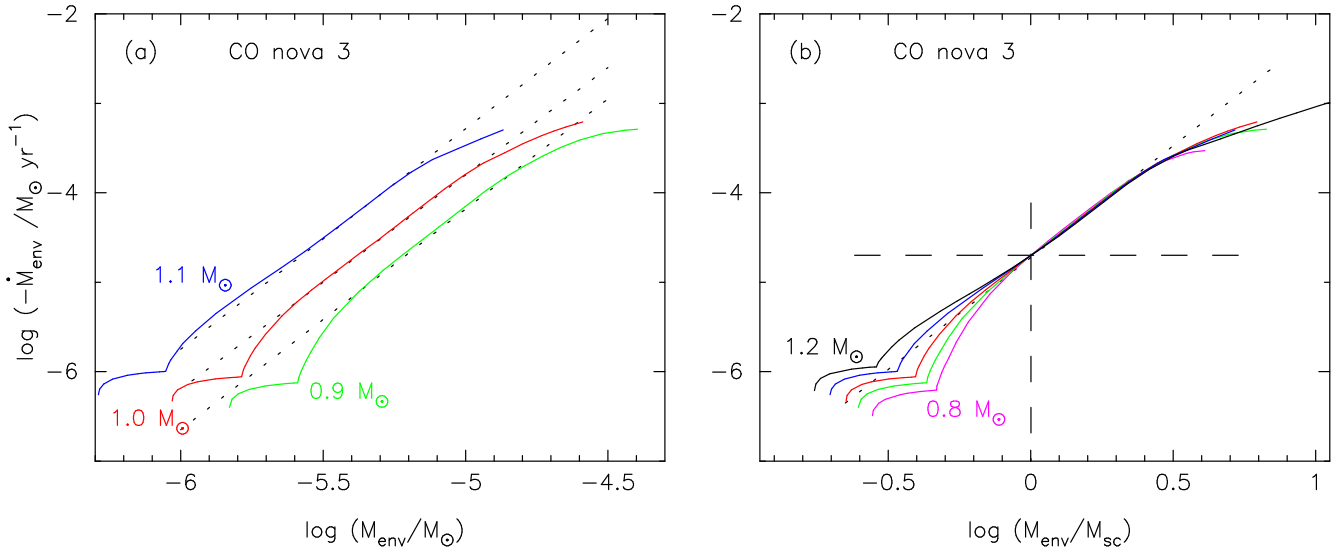}
%\plotone{dmdt_env_mass_scaling_relation_x45z02c15o20_no4.epsi}
%\plotfiddle{evolution1.ps}{5.0cm}{270}{0.4}{0.4}{-170}{220}
\caption{
(a) The envelope mass decreasing rate versus hydrogen-rich envelope mass of
optically thick wind (or static) solutions for three WD masses 
with the chemical composition of CO nova 3 (CO3).
The solid blue, red, and green lines denote $1.1~M_\sun$, $1.0~M_\sun$,
and $0.9~M_\sun$  WD models, respectively.  The break of each line
corresponds to the $M_{\rm cr}$ at which optically thick winds stop.
The dotted lines indicate the global trends of 
$\dot M_{\rm env}$-$M_{\rm env}$ relations.
(b) The ordinate is the same, but
the horizontal axis is scaled by each scaling mass $M_{\rm sc}$,
where the scaling mass is defined by each envelope mass having 
$\log (-\dot M_{\rm env}/M_\sun{\rm ~yr}^{-1}) = -4.7$.  
Two other WD mass models are added, i.e., $0.8~M_\sun$ (magenta) and
$1.2~M_\sun$ (black).  The horizontal dashed line denotes the envelope mass
decreasing rate of $\log (-\dot M_{\rm env}/M_\sun{\rm ~yr}^{-1}) = -4.7$
while the vertical dashed line indicates each scaling mass.  
\label{dmdt_env_mass_scaling_relation_x45z02c15o20}}
\end{figure*}

\citet{kat94h} calculated envelope solutions of wind mass-loss 
for various WD masses (ranging from 
$0.5~M_\sun$ to $1.38~M_\sun$) and chemical compositions.
They provide the wind mass-loss rate 
$\dot M_{\rm wind}$, photospheric temperature $T_{\rm ph}$, 
velocity $v_{\rm ph}$, and radius $R_{\rm ph}$ for a specific
envelope mass $M_{\rm env}$ and WD mass $M_{\rm WD}$.
 \citet{hac06kb, hac10k} calculated 
the nova optical and infrared (IR) light curves 
based mainly on the free-free emission model of winds.
We plot such examples in Figures
\ref{all_mass_v1668_cyg_x45z02c15o20_calib_universal_linear_no3},
\ref{dmdt_ff-flux_env_mag_relation_x45z02c15o20}, 
\ref{dmdt_env_mass_scaling_relation_x45z02c15o20}, and 
\ref{ff_flux_mag_mcr_one_fig_x45z02c15o20}
for the chemical composition of typical CO novae, i.e.,
CO nova 3 \citep[CO3;][]{hac16k}, each element of which is
$(X,Y,Z,X_{\rm C},X_{\rm O})= (0.45, 0.18, 0.02, 0.15, 0.20)$ by 
mass weight.

%Fig.2 
%\placefigure{dmdt_ff-flux_env_mag_relation_x45z02c15o20}

\subsection{Brightnesses of Optical/IR Light Curves}
\label{brightness_light_curve}
 \citet{hac06kb, hac10k} calculated the nova optical/IR light curves 
based mainly on the free-free emission model of winds.
Their free-free emission flux is approximately calculated from
\begin {equation}
F_\nu (t: M_{\rm WD}) = 
A_{\rm ff} ~{{\dot M^2_{\rm wind}} \over {v^2_{\rm ph} R_{\rm ph}}},
\label{ff_flux_wd}
\end{equation}
where they assumed $A_{\rm ff}$ to be a constant among various novae.
The effects of the electron temperature, 
ionization degree, and chemical composition on the free-free flux
are included through the envelope solutions, i.e.,
the wind mass-loss rate $\dot M_{\rm wind}$, 
photospheric velocity $v_{\rm ph}$, and photospheric radius $R_{\rm ph}$.
%because the free-free flux is mainly determined by the physical quantities
%near the photosphere.  
%If we specify $A_{\rm ff}$, we calculate the absolute $V$ magnitude.
\citet{hac10k, hac16k} determined $A_{\rm ff}$ by fitting the $0.98~M_\sun$
WD (CO3) model light curve with the V1668~Cyg light curves, 
where the distance modulus in $V$ band
$\mu_V\equiv (m-M)_V= 14.6$ as shown in Figure 
\ref{all_mass_v1668_cyg_x45z02c15o20_calib_universal_linear_no3}.
%We determined $A_{\rm ff}$ by fitting our model
%$0.98~M_\sun$ WD (CO3) light curve with the V1668~Cyg observation.

Figure \ref{all_mass_v1668_cyg_x45z02c15o20_calib_universal_linear_no3}
shows our free-free emission model light curves for 0.55, 0.60, 0.65,
0.70, 0.75, 0.80, 0.85, 0.90, 0.95, 1.0, 1.05, 1.1, 1.15, and $1.2 ~M_\sun$
WDs (CO3).  All the model light curves overlap well with each other in
the $(t/f_{\rm s})$-$(M_V - 2.5\log f_{\rm s})$ plane.
\citet{hac06kb} called this property the universal decline law.
Here, the timescaling factor
$f_{\rm s}$ is measured against the timescale of V1668~Cyg.
The light curve of this nova is well reproduced 
with a $0.98~M_\sun$ WD (CO3) model
(solid red line).  See \citet{hac16k, hac19ka, hac19kb} 
for the model light curve fitting of V1668~Cyg.

To deeply understand the physics of nova light curves,
we break the scaling process shown in Figure
\ref{all_mass_v1668_cyg_x45z02c15o20_calib_universal_linear_no3}
into two steps.  We plot the first step
in Figure \ref{dmdt_ff-flux_env_mag_relation_x45z02c15o20}(a)(b)
for three WD mass models of $0.9~M_\sun$ (green), $1.0~M_\sun$ (red),
and $1.1~M_\sun$ (blue),
that is, the free-free flux parameter
$\dot M^2_{\rm wind} / (v^2_{\rm ph} R_{\rm ph})$
against $\dot M_{\rm wind}$.
Then, we plot the second step in Figure 
\ref{dmdt_ff-flux_env_mag_relation_x45z02c15o20}(c)(d),
that is, the wind mass-loss rate $\dot M_{\rm wind}$ against the real
time $t$ and normalized time $t/f_{\rm s}$, respectively.
The third step is the combination of the first and
second steps, which is plotted in Figure
\ref{dmdt_ff-flux_env_mag_relation_x45z02c15o20}(e)(f).  
Here, we add a $0.98~M_\sun$ WD (CO3) model (black line).
Note that the ordinate in panel (e) 
is $M_V[t]-2.5\log f_{\rm s}$ because the abscissa is $t$ while
the ordinate in panel (f) is $M'_V[t/f_{\rm s}]
(= M_V[t/f_{\rm s}]-2.5\log f_{\rm s})$.  

The right column, Figure 
\ref{dmdt_ff-flux_env_mag_relation_x45z02c15o20}(b)(d)(f),
are the same as those in the left column,
but in the normalized timescale, $t/f_{\rm s}$.
Note that Figure \ref{dmdt_ff-flux_env_mag_relation_x45z02c15o20}(a)
and \ref{dmdt_ff-flux_env_mag_relation_x45z02c15o20}(b) are essentially
the same because the time does not explicitly appear. 
Combining Figure \ref{dmdt_ff-flux_env_mag_relation_x45z02c15o20}(b) with
\ref{dmdt_ff-flux_env_mag_relation_x45z02c15o20}(d), we obtain
Figure \ref{dmdt_ff-flux_env_mag_relation_x45z02c15o20}(f).
The proportionality constant $A_{\rm ff}$
is determined by fitting the $0.98~M_\sun$ WD model light curve with
the V1668~Cyg light curve.  Figure
\ref{dmdt_ff-flux_env_mag_relation_x45z02c15o20}(f) is essentially the same
as Figure \ref{all_mass_v1668_cyg_x45z02c15o20_calib_universal_linear_no3}.

The overlapping of envelope solutions in Figure
\ref{dmdt_ff-flux_env_mag_relation_x45z02c15o20}(a)(b)
directly means that, when the wind mass-loss rates are the same,
the fluxes of free-free emission are the same irrespective of
the WD mass and chemical composition.  
We express this with a function of $F_\nu(\dot M_{\rm wind})$.
This expression does not explicitly include the WD mass but 
$F_\nu(t: M_{\rm WD})$ of Equation (\ref{ff_flux_wd})
depends on the WD mass through the wind mass-loss rate 
$\dot M_{\rm wind}$ as shown in Figure
\ref{dmdt_ff-flux_env_mag_relation_x45z02c15o20}(c).
%It should be noted that the three lines in Figure
%\ref{dmdt_ff-flux_env_mag_relation_x45z02c15o20}(a)
%begin to diverge near the upper-end of the lines.
%We suppose that this diverge is due to the effect of
%numerical convergence of iterations as mentioned below in Section
%\ref{timescale_light_curves}.  We regard the true solutions to extend
%straight approximately along the dotted line, although we will not use
%this dotted line.

After the optical maximum,
the free-free flux decreases as the wind mass-loss rate drops.
This relation, $F_\nu(\dot M_{\rm wind})$, 
is common among the various WD masses and chemical compositions.  
In the real timescale of Figure
\ref{dmdt_ff-flux_env_mag_relation_x45z02c15o20}(c) and 
\ref{dmdt_ff-flux_env_mag_relation_x45z02c15o20}(e), however, the wind
mass-loss rate and flux decrease more rapidly in more massive WDs.  
To clarify the difference in the timescale between the $1.0~M_\sun$ and
$1.1~M_\sun$ WDs, we designate their times $t$ and $t'$, respectively.
For example, we plot a relation between $t$ and $t'$ 
at the vertical dashed line of 
$\log (\dot M_{\rm wind} / M_\sun~{\rm yr}^{-1})= -5.2$ in Figure
\ref{dmdt_ff-flux_env_mag_relation_x45z02c15o20}(a), which 
corresponds to the horizontal dashed line in Figure 
\ref{dmdt_ff-flux_env_mag_relation_x45z02c15o20}(c). 
The two dotted lines show the positions of $t$ and $t'$.
This $t$ and $t'$ relation similarly holds for the horizontal
dash-dotted lines of the free-free flux parameter 
$\dot M^2_{\rm wind} / (v^2_{\rm ph} R_{\rm ph})$
and $M_V-2.5\log f_{\rm s}$ in Figure
\ref{dmdt_ff-flux_env_mag_relation_x45z02c15o20}(a) and 
\ref{dmdt_ff-flux_env_mag_relation_x45z02c15o20}(e).

%Fig.3 
%\placefigure{dmdt_env_mass_scaling_relation_x45z02c15o20}

We formulate the conversion from $(t, F(t))$ to $(t', F'(t'))$ by
the time-normalization of $t'= t / f_{\rm s}$, that is,
\begin {equation}
F'(t') = f_{\rm s} F(t/f_{\rm s}).
\label{flux_conversion}
\end{equation}
This simply means that, if the timescale is squeezed by ten times
($t'=t/10$), the squeezed flux becomes larger ($F'= 10~F$) because of
energy conservation, i.e., 
$\Delta E= F'(t') \Delta t' 
= f_{\rm s} F(t') \Delta t'
= F(t/f_{\rm s}) \Delta t$.
In other words, we observe the same outburst with two different 
temporal scalings, $t$ and $t'$.  Then, the flux is different
between these two systems, $F(t)$ and $F'(t')$, while the energy
is the same, $\Delta E$, during the same intrinsic time-interval,
$\Delta t' = \Delta t /f_{\rm s}$, at the same intrinsic time, 
$t'= t/f_{\rm s}$.

From Figure \ref{dmdt_ff-flux_env_mag_relation_x45z02c15o20}(a)(c),
for example, we obtain
\begin{eqnarray}
F(t:1.0~M_\sun) & = & A_{\rm ff} ~{{\dot M^2_{\rm wind}} 
\over{v^2_{\rm ph} R_{\rm ph}}} = F'(t': 1.1~M_\sun) \cr\cr  
& = & f_{\rm s} F(t/f_{\rm s}: 1.1~M_\sun).
\label{free-free_flux_conversion}
\end{eqnarray}
Then, we convert the flux of Equation (\ref{free-free_flux_conversion})
to the absolute $V$ magnitude as
\begin{eqnarray}
& & M_V(t: 1.0~M_\sun) = M'_V(t': 1.1 ~M_\sun) \cr
&=& M_V(t/f_{\rm s}: 1.1 ~M_\sun) - 2.5 \log f_{\rm s},
\label{absolute_v_mag_scaling}
\end{eqnarray}
where $M_V$ is the absolute $V$ magnitude of the free-free emission
light curve.  The same equation holds for the $0.9~M_\sun$ 
and $1.0~M_\sun$ WDs.
%Here, the conversion from $(t, F(t))$ to $(t', F'(t'))$
%corresponds to the conversion from $(t, M_V(t))$ to $(t', M'_V(t'))$.
Thus, we derive Equation (\ref{time-stretching_general}).

\subsection{Timescales of Optical/IR Light Curves}
\label{timescale_light_curves}
The hydrogen-rich envelope mass decreases with time.  We plot 
$-\dot M_{\rm env}$ against $M_{\rm env}$ for the three WD masses
of $1.1~M_\sun$ (blue), $1.0~M_\sun$ (red), and $0.9~M_\sun$ (green)
in Figure \ref{dmdt_env_mass_scaling_relation_x45z02c15o20}(a).
Here, the decreasing rate of the envelope mass is the summation of
the wind mass-loss rate and mass-decreasing rate by nuclear burning, i.e.,
\begin{equation}
-\dot M_{\rm env} = \dot M_{\rm wind} + \dot M_{\rm nuc}.
\label{env_decrease_eq}
\end{equation}
Time goes on from the upper-right to lower-left of each line. 
The decrease in the brightness corresponds to the same color line in Figure
\ref{dmdt_ff-flux_env_mag_relation_x45z02c15o20}(e)(f).
The starting point is somewhere on the line 
at $M_{\rm env}= M_{\rm env,0}= M_{\rm ig}$,
which corresponds to the maximum brightness of a nova outburst.
Here, we specify the envelope mass at the maximum brightness by
$M_{\rm env,0}$ which we regard to be equal to the ignition mass, 
$M_{\rm ig}$.
The ignition mass is defined by the hydrogen-rich envelope mass
at the start of hydrogen burning. 
The wind mass-loss rate $\dot M_{\rm wind}$ decreases 
with decreasing envelope mass $M_{\rm env}$, and finally vanishes
when the envelope mass reaches the critical envelope mass 
$M_{\rm cr}$ required to drive a wind.  After that, the envelope mass
decreases slowly by nuclear burning.  The nova ends when the nuclear burning
extinguishes at the bottom of each line.  The sudden flattening 
on each line corresponds to the end of the wind phase. 

If we normalize the envelope mass by each scaling mass, 
$M_{\rm env}/M_{\rm sc}$, these three lines almost perfectly
overlap each other for $\log (-\dot M_{\rm env}/M_\sun{\rm ~yr}^{-1}) \ge
-4.7$ as depicted by the horizontal dashed line as shown in Figure
\ref{dmdt_env_mass_scaling_relation_x45z02c15o20}(b).
Here, we define the scaling mass by each envelope mass having 
$\log (-\dot M_{\rm env}/M_\sun{\rm ~yr}^{-1}) = -4.7$ denoted by the
vertical dashed line in Figure
\ref{dmdt_env_mass_scaling_relation_x45z02c15o20}(b).
We add other two WD masses, $0.8~M_\sun$ (magenta) and $1.2~M_\sun$ (black).
We show only five WD masses in this figure, but 
obtained the similar tendency of envelope solutions for other
WD masses (ranging from $0.6~M_\sun$ to $1.3~M_\sun$) with the same or
different chemical compositions \citep[see, e.g., Figure 6 of][]{kat94h}.

Overlapping of the five lines means that we can express 
$-\dot M_{\rm env}(x) \approx \dot M_{\rm wind}(x)$
as a function of single parameter of $x$, i.e.,
\begin{equation}
x\equiv {{M_{\rm env}} \over {M_{\rm sc}}},
\label{definition_x_envelope_mass}
\end{equation}
independently of the WD mass or chemical composition, while
the $M_{\rm sc}$ itself depends both on the $M_{\rm WD}$ and 
chemical composition, but is almost independent of the mass accretion
rate for $\dot M_{\rm acc} \lesssim 1\times 10^{-8}~M_\sun$~yr$^{-1}$.
For a larger mass accretion rate of 
$\dot M_{\rm acc} \gtrsim 3\times 10^{-8}~M_\sun$~yr$^{-1}$,
the WD radius is slightly larger compared with that of the cold core.
As a result, the $M_{\rm sc}$ for
$\dot M_{\rm acc} \gtrsim 3\times 10^{-8}~M_\sun$~yr$^{-1}$
is slightly larger than that for the cold core of
$\dot M_{\rm acc} \lesssim 1\times 10^{-8}~M_\sun$~yr$^{-1}$.
In the present paper, however, we assume that $M_{\rm sc}$
is independent of the mass accretion rate.
 
The increase with $M_{\rm env}/M_{\rm sc}$ 
in the wind mass-loss rate seems to saturate at the upper-right
end of the three lines.  In this region, $-\dot M_{\rm env}\approx 
\dot M_{\rm wind}$ because $\dot M_{\rm nuc}\ll \dot M_{\rm wind}$. 
The numerical method adopted by \citet{kat94h}
requires convergence of numerical iterations to precisely obtain the wind
mass-loss rate.  The convergence becomes very slow or fails near the
region where the lines seem to saturate.  Therefore, we suppose that 
the true wind mass-loss rate does not saturate but increases 
along the dotted line.

Then, the elapsed time is calculated from
the decreasing rate of the envelope mass, that is,
\begin{eqnarray}
t &=& \int {{d M_{\rm env}} \over {\dot M_{\rm env}}} 
= M_{\rm sc} \int {{d (M_{\rm env}/M_{\rm sc})} 
\over {\dot M_{\rm env}}}  \cr\cr\cr
&=& {{M_{\rm sc}(M_{\rm WD})} \over {M_{\rm sc,0}}}  \tau
= f_{\rm s} \tau,
\label{time-scaling_law_wd_masses}
\end{eqnarray}
where $M_{\rm sc,0}$ is a given envelope mass 
(we adopt $M_{\rm sc,0}=0.448\times 10^{-5}~M_\sun$ later in 
Equation (\ref{timescaling_mass_ratio})) and
\begin{equation}
\tau \equiv M_{\rm sc,0}\int {{d (M_{\rm env}/M_{\rm sc})} 
\over {\dot M_{\rm env}}}
= M_{\rm sc,0} \int {{d x} \over {\dot M_{\rm env}(x)}}.
\end{equation}
Therefore, we obtain
\begin{equation}
f_{\rm s} = {{M_{\rm sc}} \over {M_{\rm sc,0}}},
\label{mcritical_timescale}
\end{equation}
from the last equality in Equation (\ref{time-scaling_law_wd_masses}).
We explicitly write $M_{\rm sc}(M_{\rm WD})$ because 
the scaling mass depends on the WD mass, i.e., a function of
$M_{\rm WD}$ for a given chemical composition, as shown later in Figure
\ref{mass_cr_wd_mass}(a).  The overlapping of lines for
$\log (-\dot M_{\rm env}/M_\sun{\rm ~yr}^{-1}) \ge -4.7$ in Figure 
\ref{dmdt_env_mass_scaling_relation_x45z02c15o20}(b) guarantees
that each timescale is proportional to $f_{\rm s}\propto M_{\rm sc}$.
The fact that the envelope mass decreasing rate $\dot M_{\rm env}(x)$
is approximately a unique function of 
$x\equiv M_{\rm env}/M_{\rm sc}$ and that
the timescale is proportional to $f_{\rm s}\propto M_{\rm sc}$
are the first important conclusions of the present paper.

%Fig.4 
%\placefigure{ff_flux_mag_mcr_one_fig_x45z02c15o20}

\begin{figure*}
\epsscale{0.9}
%%\rotate
\plotone{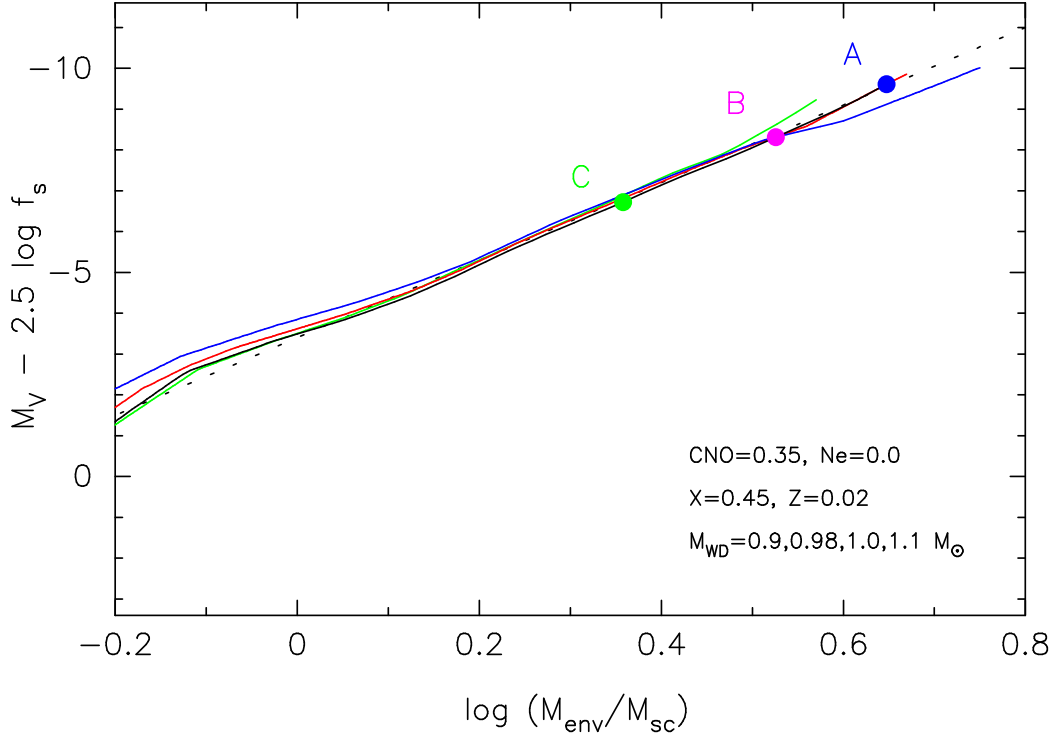}
%\plotone{ff_flux_mag_m_scaling_one_fig_x45z02c15o20_no3.epsi}
%\plotfiddle{evolution1.ps}{5.0cm}{270}{0.4}{0.4}{-170}{220}
\caption{
Free-free emission model $V$ light curves are plotted against 
$M_{\rm env}/M_{\rm sc}$ for the $0.90 ~M_\sun$ (green), 
$0.98 ~M_\sun$ (black), $1.0 ~M_\sun$ (red), and $1.1~M_\sun$ (blue) WDs.
The absolute $V$ magnitude is calibrated with the V1668~Cyg light curve
\citep{hac16k}.  The three points of A, B, and C are the same as those in
Figure \ref{all_mass_v1668_cyg_x45z02c15o20_calib_universal_linear_no3}
and specified by $x_0\equiv M_{\rm env,0}/M_{\rm sc}=4.0$, 3.1, and
2.1, respectively, on the $0.98~M_\sun$ WD model.
The dotted line represents an approximate relation of Equation
(\ref{approximate_peak_envelope_mass}).
\label{ff_flux_mag_mcr_one_fig_x45z02c15o20}}
\end{figure*}

\subsection{Universal Decline Law and Peak Brightness}
\label{universal_paek_brightness} 
Figure \ref{ff_flux_mag_mcr_one_fig_x45z02c15o20} shows
the time-normalized absolute $V$ magnitude,
$M_V - 2.5 \log f_{\rm s}$, against the scaled envelope mass,
$M_{\rm env}/M_{\rm sc}$, for $0.9~M_\sun$ (green),  
$0.98~M_\sun$ (black), $1.0~M_\sun$ (red), and $1.1~M_\sun$ (blue) WDs.
These four lines almost overlap with each other.
Therefore, $f_{\rm s} F_\nu(x)$ is approximately a unique function of 
$x= M_{\rm env}/M_{\rm sc}$ irrespective of the WD mass and chemical
composition.  This is the second important conclusion of the present paper.

This can be understood as follows: we obtain the universal decline law
in Figure \ref{dmdt_ff-flux_env_mag_relation_x45z02c15o20}(f) combining
Figure \ref{dmdt_ff-flux_env_mag_relation_x45z02c15o20}(b)
and Figure \ref{dmdt_ff-flux_env_mag_relation_x45z02c15o20}(d).
Similarly, we obtain the overlap of each line in 
Figure \ref{ff_flux_mag_mcr_one_fig_x45z02c15o20} by combining
Figure \ref{dmdt_ff-flux_env_mag_relation_x45z02c15o20}(b) and
Figure \ref{dmdt_env_mass_scaling_relation_x45z02c15o20}(b).
This is because $-\dot M_{\rm env}\approx \dot M_{\rm wind}$
for $\log (M_{\rm env}/M_{\rm sc}) = \log x \ge -0.1$ in Figure
\ref{dmdt_env_mass_scaling_relation_x45z02c15o20}(b).

The blue/green lines slightly deviate from the other lines 
for $\log (M_{\rm env}/M_{\rm sc}) \gtrsim 0.5$.
We suppose that these deviations are due to the effect of
numerical convergence of iterations as mentioned above in Section
\ref{timescale_light_curves}.
The blue line ($1.1~M_\sun$ WD) happens to be located 
below the other lines of $0.98~M_\sun$ (black) 
and $1.0~M_\sun$ (red) WDs while the green line ($0.9~M_\sun$) slightly
diverges upward at $\log (M_{\rm env}/M_{\rm sc}) \gtrsim 0.5$.
We assume that the true $M_V-2.5 \log f_{\rm s}$ for the $1.1~M_\sun$ and
$0.9~M_\sun$ WDs follow the dotted line like the other two lines.

Figure \ref{all_mass_v1668_cyg_x45z02c15o20_calib_universal_linear_no3}
shows our free-free emission model light curves for 0.55, 0.60, 0.65,
0.70, 0.75, 0.80, 0.85, 0.90, 0.95, 1.0, 1.05, 1.1, 1.15, and $1.2 ~M_\sun$
WDs (CO3).  The timescale of each WD is measured
against that of the $0.98~M_\sun$ WD (CO3), i.e.,
\begin{eqnarray}
f_{\rm s}(M_{\rm WD}) & = & {{M_{\rm sc}(M_{\rm WD})} \over
{M_{\rm sc}(0.98 ~M_\sun, {\rm CO3})}} \cr\cr
& = & {{M_{\rm sc}(M_{\rm WD})} \over
{0.448\times 10^{-5}~M_\sun}}.
\label{timescaling_mass_ratio}
\end{eqnarray}
We confirmed that the timescaling factor $f_{\rm s}$ defined by Equation 
(\ref{mcritical_timescale}) or (\ref{timescaling_mass_ratio})
is in good agreement with the timescaling factor $f_{\rm s}$ defined
directly by the light curves of the universal decline law in Figure
\ref{all_mass_v1668_cyg_x45z02c15o20_calib_universal_linear_no3}.

In short, the brightness of the universal decline law in 
Figure \ref{all_mass_v1668_cyg_x45z02c15o20_calib_universal_linear_no3}
can be specified by the two parameters, $f_{\rm s}$ and 
$\tau\equiv t/f_{\rm s}$, where $f_{\rm s}$ is related to $M_{\rm sc}$
through Equation (\ref{timescaling_mass_ratio}) and $\tau$ is related 
to $x$ with $M_V[\tau] - 2.5\log f_{\rm s} = M_V[x] - 2.5 \log f_{\rm s}$
in Figure \ref{dmdt_ff-flux_env_mag_relation_x45z02c15o20}(f) and
Figure \ref{ff_flux_mag_mcr_one_fig_x45z02c15o20}.
Thus, the two parameter set of $(f_{\rm s}, \tau)$ is equivalent to
the set of $(M_{\rm sc}, x)$.

In Figure \ref{all_mass_v1668_cyg_x45z02c15o20_calib_universal_linear_no3},
the optical peak of V1668~Cyg corresponds to point B.
Point A (C) indicates a much brighter (fainter) nova.
These points A, B, and C are also plotted in
Figure \ref{dmdt_ff-flux_env_mag_relation_x45z02c15o20}(e)(f)
against the real time $t$ and normalized time $\tau=t/f_{\rm s}$,
respectively.
Points A, B, and C correspond to different initial envelope masses
as shown in Figure \ref{ff_flux_mag_mcr_one_fig_x45z02c15o20}.
Point A, the brightest one, has $M_{V, \rm max}= -9.7$ and 
$M_{\rm env,0} = 1.8 \times 10^{-5}~M_\sun$
($x_0 \equiv M_{\rm env,0}/M_{\rm sc}=4.0$).
Point B has $M_{V, \rm max}=-8.4$ and
$M_{\rm env,0}=1.4 \times 10^{-5}~M_\sun$ ($x_0=3.1$).
Point C has $M_{V, \rm max}= -6.9$ and
$M_{\rm env,0}=0.94 \times 10^{-5}~M_\sun$ ($x_0=2.1$).
For a given WD mass ($0.98~M_\sun$), 
the brighter peak corresponds to a larger ignition mass.
Here, we approximate the initial envelope mass $M_{\rm env,0}$ by
the ignition mass $M_{\rm ig}$ of a nova outburst.
In general, the ignition mass depends on the mass accretion rate
on to the WD \citep[e.g.,][]{nom82, tow04, wol13, wol13b}.  
The lower the mass accretion rate, the larger the ignition mass
for a given WD mass.  
Thus, the nova is brighter for a smaller mass-accretion rate 
even if the WD mass is the same.

%Table 1 

\begin{deluxetable}{llll}
\tabletypesize{\scriptsize}
%%%\rotate
\tablecaption{Scaling Masses for CO Novae\tablenotemark{a}
\label{critical_mass_co2_co3_co4}}
\tablewidth{0pt}
\tablehead{
\colhead{$M_{\rm WD}$} & \colhead{$M_{\rm sc}$(CO2)} & \colhead{$M_{\rm sc}$(CO3)} & \colhead{$M_{\rm sc}$(CO4)} \\
\colhead{($M_\sun$)} & \colhead{($10^{-5}~M_\sun$)} & \colhead{($10^{-5}~M_\sun$)} & \colhead{($10^{-5}~M_\sun$)} }
\startdata
0.55 & 2.80 & 2.74 & 3.34 \\ %-4.553  0.2796E-04 & -4.562  0.2740E-04 & -4.477  0.3337E-04 \\
0.60 & 2.22 & 2.17 & 2.63 \\ %-4.655  0.2215E-04 & -4.664  0.2169E-04 & -4.581  0.2626E-04 \\
0.65 & 1.84 & 1.80 & 2.16 \\ %-4.736  0.1836E-04 & -4.746  0.1796E-04 & -4.665  0.2164E-04 \\
0.70 & 1.53 & 1.50 & 1.80 \\ %-4.815  0.1533E-04 & -4.825  0.1497E-04 & -4.745  0.1800E-04 \\
0.75 & 1.21 & 1.18 & 1.41 \\ %-4.917  0.1211E-04 & -4.929  0.1176E-04 & -4.850  0.1412E-04 \\
0.80 & 0.961 & 0.935 & 0.112 \\ %-5.017  0.9608E-05 & -5.029  0.9346E-05 & -4.953  0.1115E-04 \\
0.85 & 0.769 & 0.746 & 0.887 \\ %-5.114  0.7693E-05 & -5.127  0.7462E-05 & -5.052  0.8869E-05 \\
0.90 & 0.617 & 0.598 & 0.709 \\ %-5.210  0.6171E-05 & -5.223  0.5981E-05 & -5.149  0.7094E-05 \\
0.95 & 0.514 & 0.498 & 0.588 \\ %-5.289  0.5143E-05 & -5.303  0.4980E-05 & -5.231  0.5879E-05 \\
0.98 & 0.462 & 0.448 & ... \\ %-5.336  0.4617E-05 & -5.349  0.4479E-05 & ... \\
1.00 & 0.430 & 0.416 & 0.491 \\ %-5.367  0.4296E-05 & -5.381  0.4164E-05 & -5.309  0.4906E-05 \\
1.05 & 0.339 & 0.329 & 0.385 \\ %-5.469  0.3393E-05 & -5.483  0.3287E-05 & -5.414  0.3853E-05 \\
1.10 & 0.270 & 0.260 & 0.304 \\ %-5.569  0.2695E-05 & -5.584  0.2604E-05 & -5.517  0.3044E-05 \\
1.15 & 0.215 & 0.207 & 0.241 \\ %-5.668  0.2146E-05 & -5.683  0.2074E-05 & -5.618  0.2412E-05 \\
1.20 & 0.171 & 0.165 & 0.192 \\ %-5.768  0.1707E-05 & -5.782  0.1654E-05 & -5.716  0.1921E-05
\enddata
\tablenotetext{a}{chemical composition of the hydrogen-rich envelope
is assumed to be those of ``CO nova 2'', ``CO nova 3'', and ``CO nova 4''
in Table 2 of \citet{hac06kb}, i.e., 
$(X,Y,Z, X_{\rm C}, X_{\rm O})=(0.35, 0.33, 0.02, 0.10, 0.20)$,
$(0.45, 0.18, 0.02, 0.15, 0.20)$, and $(0.55, 0.23, 0.02, 0.10, 0.10)$,
respectively.
}
\end{deluxetable}

%Table 2 

\begin{deluxetable}{lll}
\tabletypesize{\scriptsize}
%%%\rotate
\tablecaption{Scaling Masses for Neon Novae\tablenotemark{a}
\label{critical_mass_ne2_ne3}}
\tablewidth{0pt}
\tablehead{
\colhead{$M_{\rm WD}$} & \colhead{$M_{\rm sc}$(Ne2)} & \colhead{$M_{\rm sc}$(Ne3)} \\
\colhead{($M_\sun$)} & \colhead{($10^{-5}~M_\sun$)} & \colhead{($10^{-5}~M_\sun$)} }
\startdata
%0.55 & -4.440         0.3633E-04 & -4.297         0.5041E-04 \\
%0.60 & -4.547         0.2839E-04 & -4.427         0.3743E-04 \\
%0.65 & ... & -4.514         0.3061E-04 \\
0.70 & 2.10 & 2.73 \\ %-4.72 -4.679  0.2095E-04 & -4.60 -4.564  0.2731E-04 \\
0.75 & 1.64 & 2.13 \\ %-4.785  0.1640E-04 & -4.671  0.2133E-04 \\
0.80 & 1.30 & 1.68 \\ %-4.887  0.1296E-04 & -4.776  0.1676E-04 \\
0.85 & 1.03 & 1.33 \\ %-4.986  0.1032E-04 & -4.877  0.1329E-04 \\
0.90 & 0.823 & 1.06 \\ %-5.085  0.8230E-05 & -4.976  0.1058E-04 \\
0.95 & 0.685 & 0.876 \\ %-5.165  0.6845E-05 & -5.058  0.8757E-05 \\
1.00 & 0.568 & 0.724 \\ %-5.246  0.5680E-05 & -5.140  0.7243E-05 \\
1.05 & 0.448 & 0.568 \\ %-5.349  0.4477E-05 & -5.246  0.5676E-05 \\
1.10 & 0.353 & 0.447 \\ %-5.452  0.3532E-05 & -5.350  0.4466E-05 \\
1.15 & 0.280 & 0.353 \\ %-5.552  0.2803E-05 & -5.453  0.3525E-05 \\
1.20 & 0.223 & 0.280 \\ %-5.652  0.2229E-05 & -5.553  0.2799E-05 \\
1.25 & 0.163 & 0.204 \\ %-5.788  0.1631E-05 & -5.690  0.2040E-05 \\
1.30 & 0.113 & 0.141 \\ %-5.946  0.1132E-05 & -5.851  0.1408E-05 \\
1.33 & 0.0810 & 0.100 \\ %-6.092  0.8096E-06 & -6.000  0.1001E-05 \\
1.35 & 0.0603 & 0.0743 %%%-6.220  0.6025E-06 & -6.129  0.7430E-06
\enddata
\tablenotetext{a}{chemical composition of the hydrogen-rich envelope
is assumed to be those of ``Ne nova 2'' and ``Ne nova 3'' 
in Table 2 of \citet{hac06kb}, i.e., 
$(X,Y,Z, X_{\rm O}, X_{\rm Ne})=(0.55, 0.30, 0.02, 0.10, 0.03)$
and $(0.55, 0.37, 0.02, 0.03, 0.03)$, respectively.}
\end{deluxetable}

%Table 3 

\begin{deluxetable}{ll}
\tabletypesize{\scriptsize}
%%%\rotate
\tablecaption{Scaling Masses for Solar abundance\tablenotemark{a}
\label{critical_mass_solar}}
\tablewidth{0pt}
\tablehead{
\colhead{$M_{\rm WD}$} & \colhead{$M_{\rm sc}$} \\
\colhead{($M_\sun$)} & \colhead{($10^{-5}~M_\sun$)} }
\startdata
0.55 & 7.07 \\ %-4.150  0.7072E-04 \\
0.60 & 5.88 \\ %-4.230  0.5883E-04 \\
0.65 & 4.29 \\ %-4.368  0.4286E-04 \\
0.70 & 3.35 \\ %-4.475  0.3346E-04 \\
0.75 & 2.66 \\ %-4.576  0.2656E-04 \\
0.80 & 2.22 \\ %-4.653  0.2224E-04 \\
0.85 & 1.76 \\ %-4.755  0.1756E-04 \\
0.90 & 1.39 \\ %-4.856  0.1393E-04 \\
0.95 & 1.15 \\ %-4.941  0.1146E-04 \\
1.00 & 0.948 \\ %-5.034 -5.023  0.9476E-05 \\
1.05 & 0.739 \\ %-5.131  0.7391E-05 \\
1.10 & 0.581 \\ %-5.236  0.5809E-05 \\
1.15 & 0.456 \\ %-5.341  0.4558E-05 \\
1.20 & 0.360 \\ %-5.444  0.3601E-05 \\
1.25 & 0.261 \\ %-5.584  0.2608E-05 \\
1.30 & 0.180 \\ %-5.76 -5.746  0.1795E-05 \\
1.33 & 0.127 \\ %-5.896  0.1269E-05 \\
1.35 & 0.0940 \\ %-6.027  0.9399E-06
\enddata
\tablenotetext{a}{chemical composition of the envelope is assumed
to be that of ``Solar'' in Table 2 of \citet{hac06kb}, i.e.,
$(X,Y,Z)=(0.70, 0.28, 0.02)$.}
\end{deluxetable}

\section{Approximate Analytic Relations between Various Physical Quantities}
\label{approximate_relation}
The peak brightness of a nova is calculated from the initial envelope 
mass, $M_{\rm env,0}$, as shown in Figure 
\ref{ff_flux_mag_mcr_one_fig_x45z02c15o20}.
We make an approximate analytic relation for this (black dotted line), i.e.,
\begin{equation}
M_V - 2.5\log f_{\rm s} = -9.5 ~\log \left({{M_{\rm env}}
\over {M_{\rm sc}}}\right) - 3.73.
\label{approximate_peak_envelope_mass}
\end{equation}
We tabulate $M_{\rm sc}$ for each
$M_{\rm WD}$ in Tables \ref{critical_mass_co2_co3_co4} (CO novae),
\ref{critical_mass_ne2_ne3} (Ne novae), and 
\ref{critical_mass_solar} (Solar abundance).
Then, the 1-mag, 2-mag, and 3-mag decays from the peak are defined along 
this approximate line, that is,
\begin{equation}
1 = 9.5 ~\log \left( {{M_{\rm env,0}} \over {M_{\rm env,1}}}\right),
\label{approximate_peak_1mag_decay}
\end{equation}
\begin{equation}
2 = 9.5 ~\log \left( {{M_{\rm env,0}} \over {M_{\rm env,2}}}\right),
\label{approximate_peak_2mag_decay}
\end{equation}
and
\begin{equation}
3 = 9.5 ~\log \left({{M_{\rm env,0}} \over {M_{\rm env,3}}}\right),
\label{approximate_peak_3mag_decay}
\end{equation}
where $M_{\rm env,0}$ is the envelope mass at the peak,
$M_{\rm env,1}$ at the 1-mag decay, 
$M_{\rm env,2}$ at the 2-mag decay, and $M_{\rm env,3}$ at the 3-mag
decay from the peak.  We regard $M_{\rm env,0}$ to be the same as
the ignition mass.  We derive
\begin{equation}
M_{\rm env,1} = 0.785 M_{\rm env,0},
\label{approximate_peak_1mag_decay_linear}
\end{equation}
\begin{equation}
M_{\rm env,2} = 0.616 M_{\rm env,0},
\label{approximate_peak_2mag_decay_linear}
\end{equation}
and
\begin{equation}
M_{\rm env,3}= 0.483 M_{\rm env,0}.
\label{approximate_peak_3mag_decay_linear}
\end{equation}
This means that the brightness drops by 1, 2, and 3 mag 
when the envelope mass is lost in the wind and decreases to 0.785, 0.616, 
and 0.483 times the initial envelope mass, respectively.
This property is independent of
the WD mass or chemical composition. Thus, we use the envelope mass 
that characterizes the evolution of a nova outburst instead of 
the time since the optical/IR maximum. 

Using the above property, we can rewrite the $t_1$, $t_2$, and $t_3$ times
as a function of the envelope mass.  Here, $t_1$, $t_2$, and $t_3$ times are 
\begin {equation}
t_1 = \int_{M_{\rm env,0}}^{M_{\rm env,1}} 
{{d M_{\rm env}} \over {\dot M_{\rm env}}},
\end{equation}
\begin {equation}
t_2 = \int_{M_{\rm env,0}}^{M_{\rm env,2}} 
{{d M_{\rm env}} \over {\dot M_{\rm env}}},
\end{equation}
and
\begin {equation}
t_3 = \int_{M_{\rm env,0}}^{M_{\rm env,3}} 
{{d M_{\rm env}} \over {\dot M_{\rm env}}}.
\end{equation}
We analytically approximate the wind mass loss rate 
by the dotted line in Figure
\ref{dmdt_env_mass_scaling_relation_x45z02c15o20}(b), that is,
\begin {equation}
\log \left( {{-\dot M_{\rm env}} \over
{M_\sun~{\rm yr}^{-1}}} \right) = 2.5 \log \left({{M_{\rm env}}
\over {M_{\rm sc}}}\right) - 4.7.
\label{approximate_wind_mass_loss_rate_logarithm}
\end{equation}
This equation can be rewritten as
\begin {equation}
-\dot M_{\rm env} = f_{\rm s} C_{\rm wind} \left({{M_{\rm env}}
\over {M_{\rm sc}}}\right)^{2.5},
\label{approximate_wind_mass_loss_rate}
\end{equation}
together with the proportionality constant of 
\begin {equation}
C_{\rm wind}= 10^{-4.7} M_\sun~{\rm yr}^{-1}. 
\label{constant_wind_mass_loss_rate}
\end{equation}
Using Equation (\ref{approximate_wind_mass_loss_rate}) 
together with the $x$-parameter,
$x\equiv M_{\rm env}/M_{\rm sc}$, we derive the elapsed time from
the optical peak, i.e.,
\begin {equation}
t = {M_{\rm sc} \over {C_{\rm wind}}}  \int_{x}^{x_0} 
{{d x} \over {x^{2.5}}} = {M_{\rm sc} \over 
{1.5 C_{\rm wind}}}(x^{-1.5} - x_0^{-1.5}),
\label{elapse_time}
\end{equation}
and
\begin {equation}
\tau\equiv t/f_{\rm s} 
= 54.7 (x^{-1.5} - x_0^{-1.5}) {\rm ~days}.
\label{elapse_time_tau}
\end{equation}
Then, the $t_1$ time is calculated from 
\begin {eqnarray}
t_1 &=& {M_{\rm sc} \over {C_{\rm wind}}}  \int_{x_1}^{x_0} 
{{d x} \over {x^{2.5}}} = {M_{\rm sc} \over 
{1.5 C_{\rm wind}}}(x_1^{-1.5} - x_0^{-1.5}) \cr\cr\cr
&=& {M_{\rm sc} \over {1.5 f_{\rm s} C_{\rm wind}} 
x_0^{1.5}}(0.785^{-1.5} - 1) \cr\cr\cr
&=&  {0.438 M_{\rm sc} \over {1.5 C_{\rm wind}} x_0^{1.5}},
\label{t1_time}
\end{eqnarray}
and the $t_2$ time is 
\begin {eqnarray}
t_2 &=& {M_{\rm sc} \over {1.5 f_{\rm s} C_{\rm wind}} 
x_0^{1.5}}(0.616^{-1.5} - 1) \cr\cr\cr
&=&  {1.068 M_{\rm sc} \over {1.5 C_{\rm wind}} x_0^{1.5}},
\label{t2_time}
\end{eqnarray}
and the $t_3$ time is 
\begin {eqnarray}
t_3 &=& {M_{\rm sc} \over {1.5 C_{\rm wind}} 
x_0^{1.5}} (0.483^{-1.5} -1) \cr\cr\cr
&=& {1.979 M_{\rm sc} \over {1.5 C_{\rm wind}} x_0^{1.5}}.
\label{t3_time}
\end{eqnarray}
From Equations (\ref{t1_time}) and (\ref{t2_time}),
we have a simple relation between $t_1$ and $t_2$ as
\begin{equation}
t_1 = 0.41 ~t_2,
\label{t1_t2_relation}
\end{equation}
and,  from Equations (\ref{t2_time}) and (\ref{t3_time}),
we have 
\begin{equation}
t_2 = 0.54 ~t_3.
\label{t2_t3_relation}
\end{equation}
It should be noted that Equations 
(\ref{approximate_wind_mass_loss_rate_logarithm})
and (\ref{approximate_wind_mass_loss_rate}) 
are approximately valid for
$\log (-\dot M_{\rm env}/M_\sun$~yr$^{-1}) \gtrsim -4.8$
because each line begins to diverge below that rate as shown in
Figure \ref{dmdt_env_mass_scaling_relation_x45z02c15o20}(b).
Therefore, our estimates by Equations (\ref{t1_time}), 
(\ref{t2_time}), and (\ref{t3_time}) are approximately valid for
$x_0 \gtrsim 1.16$, $x_0 \gtrsim 1.5$, and $x_0 \gtrsim 1.9$, respectively.
These lower bounds correspond to 
$\log (-\dot M_{\rm env}/M_\sun$~yr$^{-1}) = -4.8$ at
$M_{\rm env,1}$, $M_{\rm env,2}$ and $M_{\rm env,3}$, respectively.

We compare our approximate relations of Equations
(\ref{t2_time}) and (\ref{t3_time}) with the V1668~Cyg light curve.
If we adopt a $0.98~M_\sun$ WD (CO3) model, we have $x_0=3.1$ from Figure
\ref{ff_flux_mag_mcr_one_fig_x45z02c15o20},
and $M_{\rm sc}= 0.448\times 10^{-5}~M_\sun$ from Table
\ref{critical_mass_co2_co3_co4}. 
Then, we obtain
$t_2 = 10.7$ days, 
$t_3 = 19.8$ days, and the ratio $t_2/t_3 = 10.7/19.8 = 0.54$.  
These $t_2$ and $t_3$ values are slightly shorter than, 
but approximately consistent with,
the observation, e.g., $t_2= 12.2$~days and $t_3= 24.3$~days 
(for $V$ band) in \citet{mal79},
or $t_2= 12$~days and $t_3= 23$~days (for $V$ band) in \citet{dip81}.

\citet{hac06kb} discussed the relation between $t_2$ and $t_3$ based on
the universal decline law because it has a slope of 
$F_\nu \propto t^{-1.75}$.  They obtained the relation 
$t_3 = 1.69 t_2 + 0.69\Delta t_0$, where $\Delta t_0$ is the time from the
outburst to optical maximum.  Usually $\Delta t_0$ is short compared with
$t_2$ and $t_3$.  Then, we have $t_2 = (1/1.69) t_3 = 0.59 t_3$. 
This is approximately equivalent to the result of 
Equation (\ref{t2_t3_relation}).
\citet{hac06kb} compared their results with the observation.  For example,
\citet{cap90} obtained
$t_3 = (1.68\pm 0.08)~t_2 + (1.9 \pm 1.5)$ days for $t_3 < 80$ days,
or
$t_3 = (1.68\pm 0.04)~t_2 + (2.3 \pm 1.6)$ days for $t_3 > 80$ days.
If $t_3 \gg 3$ days, we obtain $t_2= (1/1.68) t_3 = 0.60 t_3$.
This is also approximately equivalent to Equation (\ref{t2_t3_relation}).

Recent work done by \citet{ozd18} concluded, however, that the relation
between $t_2$ and $t_3$ is not unique but different among various types
of nova light curve shapes defined by \citet{str10}.
\citet{ozd18} obtained 
$\log t_3 = 0.96\log t_2 + 0.32$ for S (smooth) -type,
$\log t_3 = 0.92\log t_2 + 0.43$ for P (plateau) -type,
$\log t_3 = 0.72\log t_2 + 0.6$ for D (dip) -type, and
$\log t_3 = 0.46\log t_2 + 1.29$ for J (jitter) -type.
The S-type relation corresponds to $t_2\approx 0.5 (t_3)^{1.04}$.
This is consistent with Equation (\ref{t2_t3_relation}).
We should note that the physical meaning of $t_2$ or $t_3$ is 
a local decline trend near the optical peak.  If a light curve has
multiple peaks, secondary maximum, oscillations, early dust blackout,
jitters, or flares, we should not apply $t_2$ or $t_3$
because $t_2$ or $t_3$ is greatly affected by such local variations. 

Not all but rather many novae broadly follow the universal decline law
\citep[e.g.,][]{hac06kb, hac07k, hac10k, hac16k, hac18kb, hac19ka, hac19kb}.
Strictly speaking, the universal decline law is well applied to
S-type light curve shape novae defined by \citet{str10}. 
Such an example is V1668~Cyg in Figure 
\ref{all_mass_v1668_cyg_x45z02c15o20_calib_universal_linear_no3}.
The other types of nova light curve shapes deviate from our model
light curves in some part.  However, their global trends of decline
can be sometimes fitted with our model light curves.
%%%\citep{hac06kb, hac10k, hac16k, hac18kb, hac19ka, hac19kb}. 
Our approximate formulae mentioned above are valid for such novae.

%Fig.5 
%\placefigure{max_t3_downes_selvelle_crit_x0}

\begin{figure*}
%%\epsscale{1.15}
%%\rotate
%%\plotone{f317.eps}
\plotone{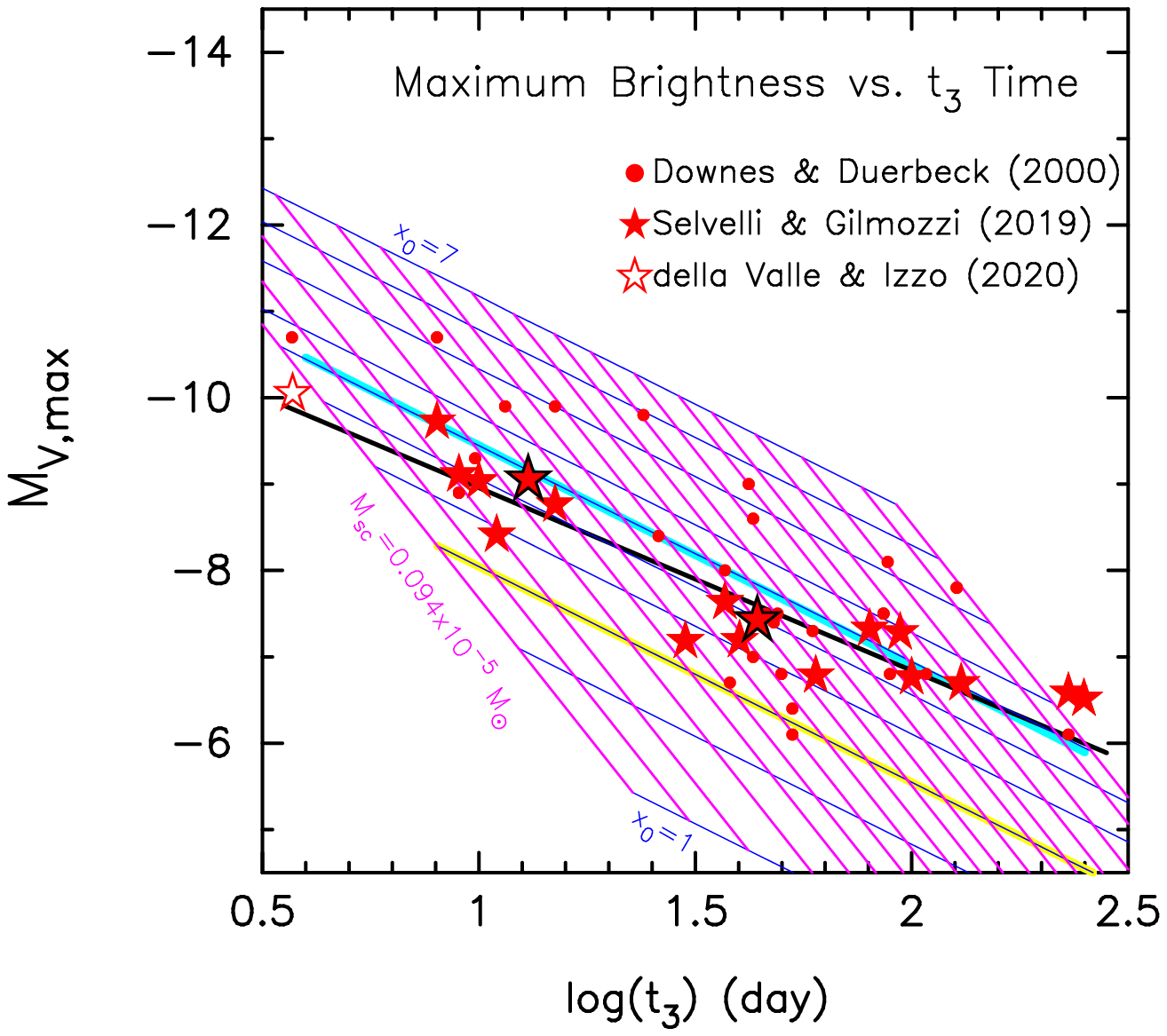}
%\plotone{max_t3_downes_selvelle_crit_x0.epsi}
%\plotfiddle{evolution1.ps}{5.0cm}{270}{0.4}{0.4}{-170}{220}
\caption{
The maximum $V$ magnitude $M_{V, \rm max}$ against the rate of decline $t_3$
for various equi-$M_{\rm sc}$ (solid magenta lines) and equi-$x_0$ 
(solid blue lines) models.  The $M_{V, \rm max}$ and $t_3$ are 
calculated from Equations (\ref{approximate_peak_envelope_mass_simple}) 
and (\ref{t3_time_simple}), respectively.
The filled red circles and stars are for galactic novae
obtained by \citet{dow00} and \citet{sel19}, respectively.
We also add an unfilled red star at the position of V1500~Cyg taken
from \citet{del20}.
The upper-left outlined red star is for GK~Per 
while the lower-right outlined red star is for V533~Her.
The thick solid cyan line represents the ``classical'' MMRD relation
defined by \citet{dow00}, i.e., $M_{V, \rm max}= -11.99 + 2.54 \log(t_3)$
while the thick solid black line represents the linear MMRD relation
defined by \citet{sel19}, i.e., $M_{V, \rm max}= -11.08 + 2.12 \log(t_3)$.
The solid magenta lines connect the same scaling mass $M_{\rm sc}$
but for different $x_0$.  
The scaling masses are, from right to left, the same as those tabulated
in Table \ref{critical_mass_solar}.
The thin solid blue lines connect the same $x_0$, i.e.,
$x_0=1$, 1.5, 2, 2.5, 3, 3.5, 4, 5, 6, and 7, from lower to upper.  
The thick yellow line corresponds to the $x_0=2$ line, above which
Equation (\ref{t3_time_simple}) is approximately valid.
The magenta lines of equi-$M_{\rm sc}$ have a slope of 6.3 while
the blue lines of equi-$x_0$ have a slope of 2.5 in the 
$(\log t_3)$-$M_{V, \rm max}$ diagram. 
\label{max_t3_downes_selvelle_crit_x0}}
\end{figure*}

\section{Theoretical MMRD Relation}
\label{theoretical_mmrd_relation}

\subsection{The MMRD Relation of the Universal Decline Law}
\label{mmrd_universal_decline}

In the previous section, we formulated the nova model light curves
for various WD masses and chemical compositions by the two parameters
of $(M_{\rm sc}, x_0)$.  In this section, we convert these two parameters
to $(t_3, M_{V, \rm max})$.
From Equation (\ref{t3_time}),
we have
\begin {eqnarray}
t_3  =  19.8 \left( {{M_{\rm sc}} \over {4.48\times 10^{-6}~M_\sun}}
\right) ~\left( {{x_0} \over {3.1}} \right)^{-1.5} {~\rm days}.
%%%& = & 19.7 ~f_{\rm s}~\left( {{x_0} \over {3.1}} \right)^{-1.5} {~\rm days}.
\label{t3_time_simple}
\end{eqnarray}
This is approximately valid for $x_0 \gtrsim 2.0$ 
as noted in Section \ref{approximate_relation}. 
Also from Equation (\ref{approximate_peak_envelope_mass}), we have
\begin{eqnarray}
M_{V,\rm max} & = &  2.5\log \left( {{M_{\rm sc}} \over 
{4.48\times 10^{-6}~M_\sun}} \right) \cr\cr
& & -9.5 \log \left({{x_0}\over {3.1}}\right) - 8.4.
%%%& = & 2.5 \log f_{\rm s} -9.5 \log \left({{x_0}\over {3.1}}\right) -8.4.
\label{approximate_peak_envelope_mass_simple}
\end{eqnarray}
This is approximately valid for $x_0 \gtrsim 1.0$ 
as shown in Figure \ref{ff_flux_mag_mcr_one_fig_x45z02c15o20}.
To summarize, we can apply Equations (\ref{t3_time_simple}) and
(\ref{approximate_peak_envelope_mass_simple}) for $x_0 \gtrsim 2.0$.

We plot equi-$M_{\rm sc}$ lines and equi-$x_0$ lines in Figure
\ref{max_t3_downes_selvelle_crit_x0}.
The solid magenta lines represent each equi-$M_{\rm sc}$ line corresponding
to the $M_{\rm sc}$ values in Table \ref{critical_mass_solar}.
The thin solid blue lines denote each equi-$x_0$ line, from bottom to top,
$x_0=1$, 1.5, 2, 2.5, 3, 3.5, 4, 5, 6, and 7.  
The thick yellow line corresponds to the $x_0=2$ line, above which
Equations (\ref{t3_time_simple}) and 
(\ref{approximate_peak_envelope_mass_simple}) are approximately valid. 

We add observational points taken from \citet{dow00} with filled red circles
and \citet{sel19} with filled red stars.
We also add an unfilled red star at the position of V1500~Cyg taken
from \citet{del20}.  For the distance to a nova,
\citet{dow00} used the expansion parallax method of nova shells 
while \citet{sel19} and \citet{del20} used the trigonometric parallaxes
of Gaia Data Release 2 (Gaia DR2).  These two data show a similar trend 
in the $(\log t_3)$-$M_{V,\rm max}$ diagram.  We also add two linear 
trend lines of $M_{V, \rm max}= -11.08 + 2.12 \log t_3$ (thick
solid black line) and $M_{V, \rm max}= -11.99 + 2.54 \log t_3$ (thick
solid cyan line) that represent linear trends derived by \citet{sel19}
and \citet{dow00}, respectively.
The upper-left outlined red star is for GK~Per 
while the lower-right outlined red star
is for V533~Her both from \citet{sel19}.
We discuss these two novae in Section \ref{discussion}.

The equi-$M_{\rm sc}$ lines have a slope of 6.3 ($=9.5/1.5$) while
the equi-$x_0$ lines have a slope of 2.5 in the $(\log t_3)$-$M_{V, \rm max}$
diagram.  The latter slope is close to 2.54 of the cyan line, the trend
MMRD line defined by \citet{dow00}, but slightly steeper than
the slope of 2.12 (thick solid black line) obtained by \citet{sel19}.
This thick black line traverses the two blue lines of $x_0= 3.0$ and 3.5
from left to right. 
The observational MMRD points (both the filled red circles and stars)  
are covered with the region between the lower bound $x_0\sim 2$ and 
the upper bound $x_0\sim 6$, the center of which is $x_0\sim 3.5$.
\citet{hac10k} examined the observational MMRD distribution
obtained by \citet[][filled red circles]{dow00} and reached a similar
conclusion \citep[see Figure 15 of][]{hac10k}.  This is because
the trend MMRD line (cyan line) obtained by \citet{dow00} 
almost overlaps with the blue $x_0= 3.5$ line.

It should be noted that the global timescale of a nova light curve
is $f_{\rm s}\propto M_{\rm sc}$ as shown in Equations
(\ref{mcritical_timescale}) and (\ref{timescaling_mass_ratio}).
On the other hand, the $t_3$ time is not a global timescale but
a local timescale only near the peak of a nova light curve.  
The $t_3$ time is proportional to $f_{\rm s}(\propto M_{\rm sc})$
but depends also on the $x_0$ as in Equation (\ref{t3_time})
or (\ref{t3_time_simple}).  This is the reason why the MMRD points
show a large scatter around the main trend in the $(\log t_3)$-$M_{V,\rm max}$
diagram.  This can be easily understood if we eliminate $M_{\rm sc}$
from Equations (\ref{t3_time_simple}) and 
(\ref{approximate_peak_envelope_mass_simple}) and obtain
\begin{equation}
M_{V,\rm max}= 2.5\log t_3 - 11.95 
- 5.75 \log \left({{x_0}\over {3.5}}\right).
\label{approximate_mmrd_relation_x0}
\end{equation}
Then, we have 
\begin{equation}
M_{V,\rm max}= 2.5\log t_3 - 11.95, {\rm ~for~}x_0=3.5.
\label{approximate_mmrd_relation_x0_76}
\end{equation}
This MMRD relation is essentially the same as that (cyan line)
obtained by \citet{dow00},
but slightly steeper than the trend MMRD line (black line) obtained
by \citet{sel19} in Figure \ref{max_t3_downes_selvelle_crit_x0}. 
The scatter from the main trend ($x_0\sim 3.0-3.5$ line) can be 
understood from the difference in $x_0$, that is, the difference
in the ignition mass.

%Fig.6 
%\placefigure{max_t3_downes_all_saio_kato2014}

\begin{figure*}
%%\epsscale{1.15}
%%\rotate
%%\plotone{f317.eps}
\plotone{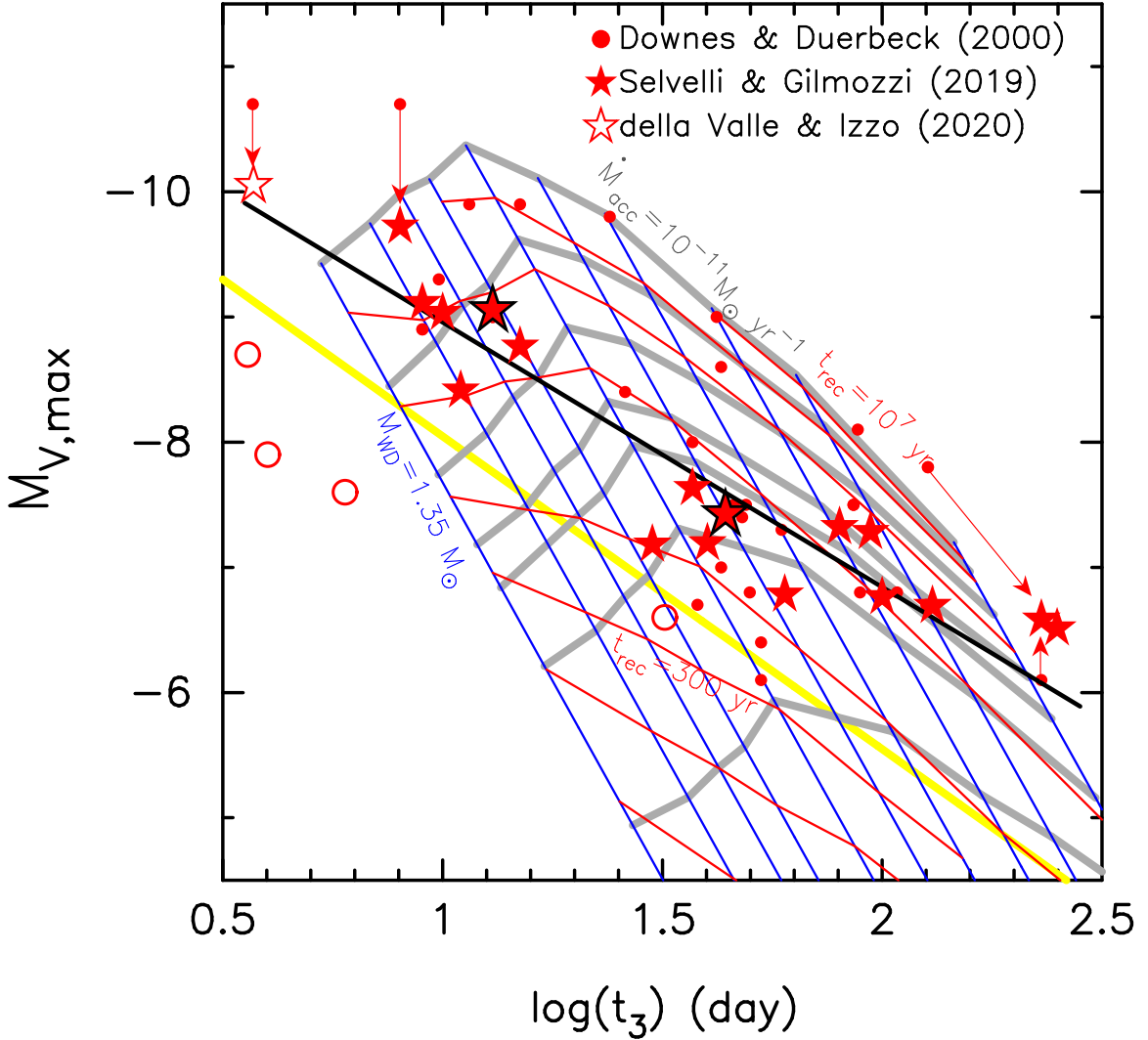}
%\plotone{max_t3_downes_all_saio_kato2014_no2.epsi}
%\plotfiddle{evolution1.ps}{5.0cm}{270}{0.4}{0.4}{-170}{220}
\caption{
Same as Figure \ref{max_t3_downes_selvelle_crit_x0}, but for 
equi-$M_{\rm WD}$, equi-$\dot M_{\rm acc}$, and equi-$t_{\rm rec}$
models of solar abundance.  
The thin solid blue lines connect the same WD mass $M_{\rm WD}$
but for different mass accretion rates $\dot M_{\rm acc}$.  
The WD masses are, from right to left, $M_{\rm WD}= 0.6$, 0.7, 0.8, 0.9,
1.0, 1.1, 1.2, 1.25, 1.3, and $1.35~M_\sun$.
The thick yellow line corresponds to the $x_0=2$ line.
The thin solid red lines connect the same recurrence time, i.e.,
$t_{\rm rec}=30$, 100, 300, 1000, 10000, $10^5$, $10^6$, and $10^7$~yr,
from lower to upper.  
The thick solid gray lines represent the same mass accretion rate,
from lower to upper, $\dot M_{\rm acc}= 3\times 10^{-8}$,
$1\times 10^{-8}$, $5\times 10^{-9}$, $3\times 10^{-9}$, 
$1\times 10^{-9}$, $1\times 10^{-10}$,
and $1\times 10^{-11}~M_\sun$~yr$^{-1}$. 
See Table \ref{mmrd_relation_kato2014} for each data.
The four unfilled red circles denote the MMRD positions of the four recurrent
novae, CI~Aql, T~CrB, U~Sco, and V745~Sco.
Other symbols and lines are the same as those in Figure 
\ref{max_t3_downes_selvelle_crit_x0}.  See text for more details.
\label{max_t3_downes_all_saio_kato2014}}
\end{figure*}

\subsection{The MMRD Relation for Novae}
\label{peak_brightness_vs_decline}
In this subsection, we obtain $(t_3, M_{V,\rm max})$ against
nova models, which are specified by 
the mass accretion rate $\dot M_{\rm acc}$ and the WD mass $M_{\rm WD}$
(or sometimes by the recurrence time $t_{\rm rec}$).
These parameters are determined by the binary nature.  
We adopt the ignition mass $M_{\rm ig}$, mass accretion rate 
$\dot M_{\rm acc}$, and recurrence time $t_{\rm rec}$ from
published data available to the authors.   

\subsubsection{Ignition mass model}
\label{ignition_mass_model}
The ignition masses have been calculated by many authors
\citep[e.g.,][]{pri95, kat14shn, hac16sk, chen19}.
However, the ignition masses can sometimes differ significantly
each other, depending not only on the model assumption but also on 
the calculation method \citep[numerical code, see, e.g.,][]{kat17palermo}.

We adopt the result of \citet{kat14shn}.
They calculated the accumulation mass $M_{\rm acc}$ and ignition mass
$M_{\rm ig}$ assuming solar abundance.  They obtained models for
$\dot M_{\rm acc} \ge 1\times 10^{-9}~M_\sun$~yr$^{-1}$.
In the present paper, we extend the mass accretion rate down to 
$\dot M_{\rm acc} \ge 1\times 10^{-11}~M_\sun$~yr$^{-1}$ taking into
account the lower mass-accretion rate limit for cataclysmic variables
\citep[e.g.,][]{kni11}.  We tabulate $M_{\rm acc}$ (third column) 
in Table \ref{mmrd_relation_kato2014} for various WD masses $M_{\rm WD}$
(first column) and mass accretion rates $\dot M_{\rm acc}$ (second column).
We assume that the envelope mass at optical maximum $M_{\rm env,0}$
is almost the same as the ignition mass $M_{\rm ig}$, i.e.,
$M_{\rm env,0}= M_{\rm ig}$.
Using the scaling mass $M_{\rm sc}$ in Table \ref{critical_mass_solar},
we calculate the ratio of $x_0\equiv M_{\rm env,0}/M_{\rm sc}= 
M_{\rm ig}/M_{\rm sc}$.
These $x_0$ are also tabulated on the fifth column 
in Table \ref{mmrd_relation_kato2014}.
%The ignition mass is not tabulated but calculated from 
%$M_{\rm ig}= x_0 \times M_{\rm sc}$ from Table \ref{mmrd_relation_kato2014}.

It should be noted that the accumulation mass 
$M_{\rm acc} = t_{\rm rec} \times \dot M_{\rm acc}$ is slightly smaller than 
the envelope mass at ignition $M_{\rm ig}= M_{\rm env,0} = x_0 \times
M_{\rm sc}$.  When the mass accretion starts, there is residual 
hydrogen-rich material on the WD.
This residual is leftover from the previous nova explosion.
The ignition mass is defined by the summation of the accreted mass and 
the envelope mass at the epoch when hydrogen burning extinguishes, i.e.,
\begin{equation}
M_{\rm ig}= M_{\rm env,0}=  M_{\rm acc} + M_{\rm env,min}. 
\label{minimum_envelope_mass}
\end{equation}
(See Figure 1 of \citet{kat14shn} for the relation 
among these envelope masses.)

For a given ignition mass,
we obtain the absolute $V$ magnitude at optical maximum 
$M_{V, \rm max}$ from Equation 
(\ref{approximate_peak_envelope_mass_simple}).
The values of $M_{V, \rm max}$ are tabulated at 6th column in 
Table \ref{mmrd_relation_kato2014}.  
We obtain the $t_3$ time from Equation (\ref{t3_time_simple})
and $t_2$ time from Equation (\ref{t2_t3_relation}).
These are also tabulated in Table \ref{mmrd_relation_kato2014}.

\subsubsection{Global trend of MMRD relation}
\label{global_trend_mmrd}
We plot these peak $V$ brightness versus rate of decline relation
in Figure \ref{max_t3_downes_all_saio_kato2014}.
Each thin solid blue line connects
the same WD mass models with different mass accretion rates.
The thick solid gray lines connect the models with the same mass
accretion rate.  These gray lines have a peak of $M_{V, \rm max}$ at
$M_{\rm WD}= 1.1 M_\sun$.  The thin red lines connect the same recurrence
period models.  

In this figure, we add the observational MMRD points, filled red circles,
filled red stars, and an unfilled red star, obtained by \citet{dow00},
\citet{sel19}, and \citet{del20}, respectively.  
We can see that the distribution of MMRD points is covered by
mass accretion rates between
$\dot M_{\rm acc}\sim 1\times 10^{-11}~M_\sun$~yr$^{-1}$
and 
$\dot M_{\rm acc}\sim 3\times 10^{-8}~M_\sun$~yr$^{-1}$ 
and centered on
$\dot M_{\rm acc}\sim 5 \times 10^{-9}~M_\sun$~yr$^{-1}$
in the longer $t_3$ region of $t_3\gtrsim 30$~days,
based on the result of \citet{kat14shn}.

The thick solid black line is a trend of MMRD distribution obtained
by \citet{sel19}, which is close to an equi-mass accretion rate line
of $\dot M_{\rm acc}\sim 5 \times 10^{-9}~M_\sun$~yr$^{-1}$ in the
longer $t_3$ region of $t_3\gtrsim 30$~days (or in the fainter region
of $M_{V, \rm max} \ge -8.0$).

On the other hand, the thick black line traverses the four gray lines of
$\dot M_{\rm acc}= 3\times 10^{-9}~M_\sun$~yr$^{-1}$,
$1\times 10^{-9}~M_\sun$~yr$^{-1}$,
$1\times 10^{-10}~M_\sun$~yr$^{-1}$, and
$1\times 10^{-11}~M_\sun$~yr$^{-1}$ 
in the brighter region of $M_{V, \rm max} \le -8.0$.
If we divide the data set (filled red stars) of \citet{sel19} into two groups
by the brightness of $M_{V, \rm max} = -8.0$, the brighter (upper-left) 
group is located in between $\dot M_{\rm acc}= 1\times 10^{-9}$
and $1\times 10^{-11}~M_\sun$~yr$^{-1}$ and the fainter (lower-right) group
is located in between $\dot M_{\rm acc}= 3\times 10^{-8}$,
and $3\times 10^{-9}~M_\sun$~yr$^{-1}$.  This difference in the mass
accretion rate broadly correspond to cataclysmic variables below/above
the period gap \citep[e.g.,][]{kni11}.  

For an MMRD point of the observed nova, we are able to broadly specify 
the WD mass and mass accretion rate.
The main trend in the MMRD diagram corresponds to the mass accretion
rate of $\dot M_{\rm acc} \sim 5\times 10^{-9}~M_\sun$~yr$^{-1}$ with
the different WD masses in the lower region of $M_{V, \rm max} \ge -8.0$.  
\citet{sel19} estimated the mass accretion rates 
from the quiescent luminosities of their data set novae
in Figure \ref{max_t3_downes_all_saio_kato2014} (filled red stars).
Their median value is 
$\dot M_{\rm WD} = 3.3\times 10^{-9}~M_\sun$~yr$^{-1}$,
which is close to our main trend value of 
$\dot M_{\rm acc}\sim 5\times 10^{-9}~M_\sun$~yr$^{-1}$.  
The scatter up to $\pm 3\sigma= \pm 1.0$ mag from the main trend line
can be attributed to the different mass accretion rates.
{\it
Thus, the global trend of an MMRD relation does indeed exist, at least,
in the fainter region of $M_{V, \rm max} \ge -8.0$,  
but its scatter is too large for it to be a precision distance indicator
of individual novae.
}

\citet{sel19} reported a correlation between $\dot M_{\rm acc}$ and
the speed class $t_3$ ($\dot M_{\rm acc}$ increasing with $t_3$ as in
their Figure 5) for their 17 nova data set and wrote ``we cannot find
a simple explanation that could account for this.''
% How, for example,
%does $t_3$, a likely proxy for the WD mass, control $\dot M_{\rm acc}$
%between outbursts?  Clearly this requires a more in-depth theoretical
%analysis. Yet, this is a useful relation that allows one to estimate
%$\dot M_{\rm acc}$ directly, and marks $t_3$ as an extremely convenient
%observable to evaluate several critical parameters of classical novae.''
This correlation ($\dot M_{\rm acc}$ increases with $t_3$)
can be easily seen up to $t_3 \lesssim 80$~days for the filled red stars
in our Figure \ref{max_t3_downes_all_saio_kato2014}. 
For $t_3 \gtrsim 80$~days, this correlation seems to be flat
except for RR~Pic and HR~Del. 
This global trend of our $\dot M_{\rm acc}$ versus 
$t_3$ relation is consistent with their Figure 5.
%However, it should be noted that this correlation is valid {\it only
%for the 17 nova data set of \citet{sel19}.}  It is obvious that this
%correlation is not correct if we include the group of recurrent novae,
%in which the mass-accretion rate is rather high but $t_3$ is very short.    

\subsubsection{MMRD positions of individual novae}
\label{individual_novae_mmrd}
Finally, we point out that there is a $-10.4$ mag cap for the maximum
$V$ brightness if we limit the mass accretion rate above 
$\dot M_{\rm acc} \ge 1\times 10^{-11} ~M_\sun$~yr$^{-1}$.  

Several novae are located outside our region of $(t_3, M_{V, \rm max})$.
They are, from left to right, V1500~Cyg (red filled circle: $t_3= 3.7$~days, 
$M_{V, \rm max}= -10.7$), CP~Pup (red filled circle: $8$~days, $-10.7$),
RR~Pic (red filled circles: $127$~days, $-7.8$), HR~Del (red filled circle:
$230$~days, $-6.1$).  The data of V1500~Cyg is revised by \citet{del20}
to be $M_{V, \rm max}= -10.05$ (unfilled red star).
The data of CP~Pup, RR~Pic, and HR~Del are revised by
\citet{sel19} to be $M_{V, \rm max}= -9.72$ and $t_3= 8$~days (CP~Pup, filled
red star), $-6.51$ and $250$~days (RR~Pic, filled red 
star), and $-6.58$ and $230$~days (HR~Del, filled red star).
These four corrections are indicated
by the red arrows in Figure \ref{max_t3_downes_all_saio_kato2014}. 

This explains that the brightest peaks of novae are fainter than
$M_V \sim -10.4$ \citep[e.g.,][]{coh85, shafter11, cao12,
shafter13, shara17, del20}.  
{\it The brightness cap ($M_V \sim -10.4$) of classical novae is constrained
by the lowest mass-accretion rate of $\dot M_{\rm acc} \sim 1\times 10^{-11}
~M_\sun$~{\rm yr}$^{-1}$. 
}

%Fig.7 
%\placefigure{mass_cr_wd_mass}

\begin{figure*}
%%\epsscale{1.15}
%%\rotate
%%\plotone{f317.eps}
\plotone{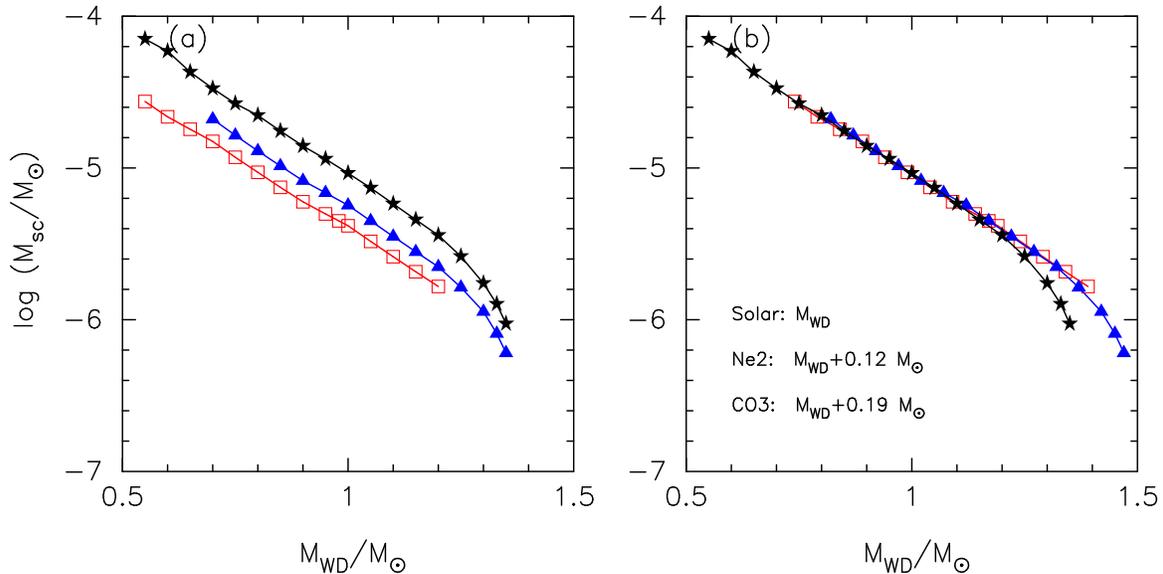}
%\plotone{mass_cr_wd_mass.epsi}
%\plotfiddle{evolution1.ps}{5.0cm}{270}{0.4}{0.4}{-170}{220}
\caption{
(a) The scaling mass $M_{\rm sc}$ 
against the WD mass $M_{\rm WD}$ for various chemical compositions.
The filled black stars connected by a black line show the models in Table
\ref{critical_mass_solar} (solar composition).
The filled blue triangles connected by a blue line denote the models in
Table \ref{critical_mass_ne2_ne3} (Ne nova 2).
The unfilled red squares connected by a red line correspond to the models
in Table \ref{critical_mass_co2_co3_co4} (CO nova 3). 
The CO3 has $X=0.45$, $Y=0.18$, $Z=0.02$, 
$X_{\rm C}=0.15$, and $X_{\rm O}=0.20$ by mass weight.
The Ne2 is composed of $X=0.55$, $Y=0.30$, $Z=0.02$,
$X_{\rm O}=0.10$, and $X_{\rm Ne}=0.03$.  
(b) Same as in panel (a), but we shift the WD masses of the chemical
composition Ne2 by $+0.12~M_\sun$, and of CO3 by $+0.19~M_\sun$ as
plotted in the figure.  These three lines overlap well until $1.2~M_\sun$.
\label{mass_cr_wd_mass}}
\end{figure*}

In the longer $t_3$ time, only two novae (HR~Del and RR~Pic) are outside
our theoretical region between $M_{\rm WD}= 0.6$ and $1.35~M_\sun$ and
between $\dot M_{\rm acc}= 3\times 10^{-8}~M_\sun$~yr$^{-1}$ and 
$1\times 10^{-11}~M_\sun$~yr$^{-1}$.  This simply means (1) that the WD mass
is smaller than $0.6~M_\sun$ or (2) that the concept of $t_2$ and $t_3$ time
should not be applied to these novae for some reasons.
%The longest $t_3$ novae in the data set of \citet{sel19} are HR~Del
%($t_3= 230\pm 4$ days) and RR~Pic ($250\pm 30$ days).  
These two novae have multiple peaks 
\citep[see, e.g., Figure 1 of ][]{hac15k}.  
The $t_2$ or $t_3$ time is a local timescale
near the peak and is greatly affected by such variations.
We should not apply $t_2$ and $t_3$ to such novae.  
\citet{hac15k} analyzed the light curves of HR~Del and RR~Pic and globally 
fitted their model light curves of $M_{\rm WD}= 0.55~M_\sun$ and 
$0.51~M_\sun$ with the observed light curves of these two novae
except the multiple peaks. 

GK~Per and V446~Her belong to the shortest $t_3$ and brightest 
$M_{V,\rm max}$ group of Selvelli \& Gilmozzi's nova data set, i.e.,
$(t_3, M_{V,\rm max})= (13, -9.05)$ and $(15, -8.76)$.
These two novae show dwarf nova outbursts in a post-nova phase,
suggesting that their mass accretion rates are now quite low,
below a few times $10^{-9}~M_\sun$~yr$^{-1}$ \citep[e.g.,][]{kni11}.
Figure \ref{max_t3_downes_all_saio_kato2014} indicates relatively
low mass accretion rates of $\sim 2\times 10^{-10}$ and
$\sim 7\times 10^{-10}~M_\sun$~yr$^{-1}$.
\citet{sel19} estimated the mass accretion rates,
$\dot M_{\rm acc}= 2\times 10^{-9}$ and $0.8\times 10^{-9}~M_\sun$~yr$^{-1}$,
respectively.  Their estimated rate of V446~Her seems
to be consistent with our theoretical value estimated from the position
in the $(\log t_3)$-$M_{V, \rm max}$ diagram while that of GK~Per 
is about ten times larger than our value.    
This suggests that, for a certain kind of novae,
the mass accretion rate during a post-nova phase
is rather high compared with that of a substantial pre-nova phase, or
that the mass accretion rate gradually decreases toward much lower
rates in a timescale of recurrence period.

CP~Pup has a lowest value of $\dot M_{\rm acc}$ in 
Selvelli \& Gilmozzi's nova data set, i.e., 
$\dot M_{\rm acc}= 0.6\times 10^{-9}~M_\sun$~yr$^{-1}$.
On the other hand, Figure \ref{max_t3_downes_all_saio_kato2014} indicates
$\dot M_{\rm acc}= (1.5-2)\times 10^{-11}~M_\sun$~yr$^{-1}$.
This is about 30 times smaller than that of Selvelli \& Gilmozzi.
\citet{schaefer10} reported that the pre-nova brightness of CP~Pup
is $\Delta m\sim 5$ mag fainter than that of a post-nova brightness.
This simply indicates that the mass-accretion rate is $\sim 100$ times
smaller than that during a post-nova phase.  If the $\dot M_{\rm acc}$
decreases toward $\dot M_{\rm acc}= 0.6\times 10^{-11}~M_\sun$~yr$^{-1}$
at the pre-nova phase, 
the average mass-accretion rate during the quiescent phase is
close to $\dot M_{\rm acc}\sim 1\times 10^{-11}~M_\sun$~yr$^{-1}$.  
This average value is consistent with our theoretical estimate.

\subsubsection{MMRD positions of recurrent novae}
\label{recurrent_novae_mmrd}
Recurrent novae (unfilled red circles) are located below the yellow line 
in Figure \ref{max_t3_downes_all_saio_kato2014}. 
Here, we plot four MMRD positions of the recurrent novae,
CI~Aql $(t_3= 32~{\rm days}, M_{V, \rm max}= -6.6)$, T~CrB $(6, -7.6)$,
U~Sco $(3.6, -8.7)$, and V745~Sco $(4, -7.9)$.  The data are taken
from Table 2 of \citet{hac18kb} except for CI~Aql.  The data of 
CI~Aql are calculated from $m_{V, \rm max}=9.0$ \citep{str10}, 
$A_V= 3.1 E(B-V)= 3.1\times 1.0= 3.1$ \citep{hac18kb},
and $d=3189^{+949}_{-315}$~pc \citep[Gaia distance,][]{schaefer18}.
We exclude RS~Oph and T~Pyx.  This is because the light curve of RS~Oph
is contaminated by the shock-heating between the ejecta and 
circumstellar matter and, as a result, the $t_3$ time does not represent
the timescale of intrinsic decline \citep[e.g.,][]{hac18kb}.
T~Pyx has multiple peaks and we should not apply the concept of $t_3$.

The recurrence periods of these recurrent novae are $\sim 10 - 80$ yr
\citep[e.g.,][]{schaefer10a}.
It is obvious that the MMRD positions of these recurrent novae are 
not consistent with the equi-recurrence period lines of $t_{\rm rec}=30$
and 100 yr in Figure \ref{max_t3_downes_all_saio_kato2014}.
We give three reasons for this exceptional case.
The first point is that the present theory cannot be applicable to 
recurrent novae because Equation (\ref{t3_time_simple}) is not valid for
$x_0\lesssim 2$ (below the yellow line) 
as mentioned in Section \ref{approximate_relation}. 
%%%We add other two points.
The second point is that the mass accretion rate is probably larger than
$\dot M_{\rm acc} \gtrsim 3\times 10^{-8}~M_\sun$~yr$^{-1}$ in recurrent
novae.  This means that the WD core is hot and its radius is larger
than that of a cold core as mentioned in Section \ref{timescale_light_curves}.
We must calculate wind solutions assuming the large WD radius depending on
the mass accretion rate.  The third point is that a hot helium layer develops
underneath a hydrogen-rich envelope in recurrent novae 
\citep[e.g.,][]{kat17shb}.  This possibly affects the thermal state of
hydrogen-rich envelope.  We must take into account at least these three
effects for recurrent novae.  The calculation is so complicated that
we leave it to near future.   

%%%xxxxxxxxxxxxxxxxxxxxxxxxx  

%Fig.8 
%\placefigure{max_t3_downes_selvelle_crit_x0_co3}

\begin{figure*}
%%\epsscale{1.15}
%%\rotate
%%\plotone{f317.eps}
\plotone{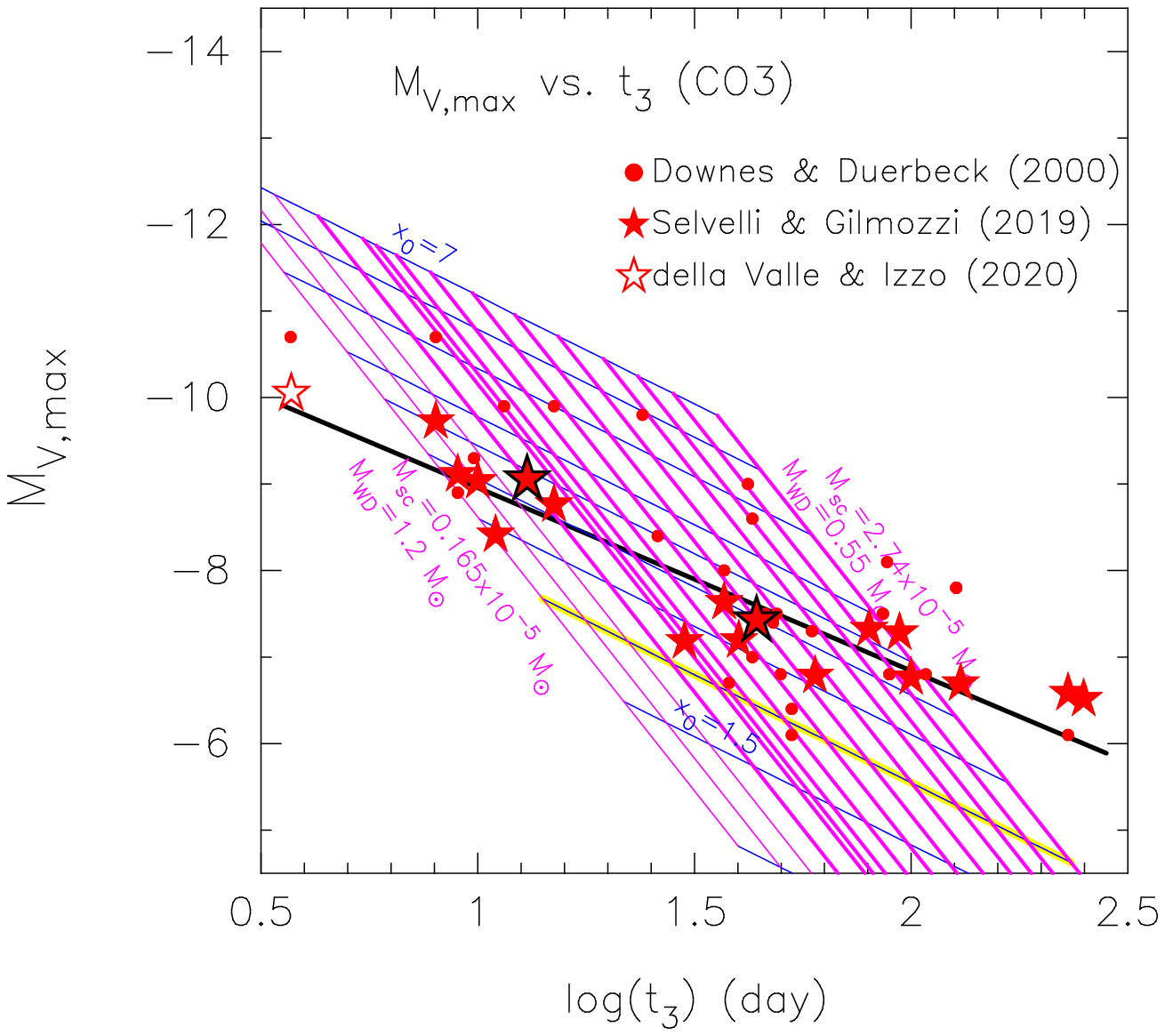}
%\plotone{max_t3_downes_selvelle_crit_x0_co3.epsi}
%\plotfiddle{evolution1.ps}{5.0cm}{270}{0.4}{0.4}{-170}{220}
\caption{
Same as Figure \ref{max_t3_downes_selvelle_crit_x0}, but for the chemical
composition of CO nova 3 (CO3).
Each scaling mass (magenta line) corresponds to the WD mass, 
from right to left, $0.55~M_\sun$
to $1.2~M_\sun$ by $0.05~M_\sun$ step except $0.98~M_\sun$,
the same as those tabulated
in Table \ref{critical_mass_co2_co3_co4}(CO3).
The thick solid magenta lines of equi-$M_{\rm sc}$ 
(or equi-$M_{\rm WD}$) correspond to
the WD mass region for CO novae ($M_{\rm WD} \le 1.05~M_\sun$).
\label{max_t3_downes_selvelle_crit_x0_co3}}
\end{figure*}

%Fig.9 
%\placefigure{max_t3_downes_selvelle_crit_x0_ne2}

\begin{figure*}
%%\epsscale{1.15}
%%\rotate
%%\plotone{f317.eps}
\plotone{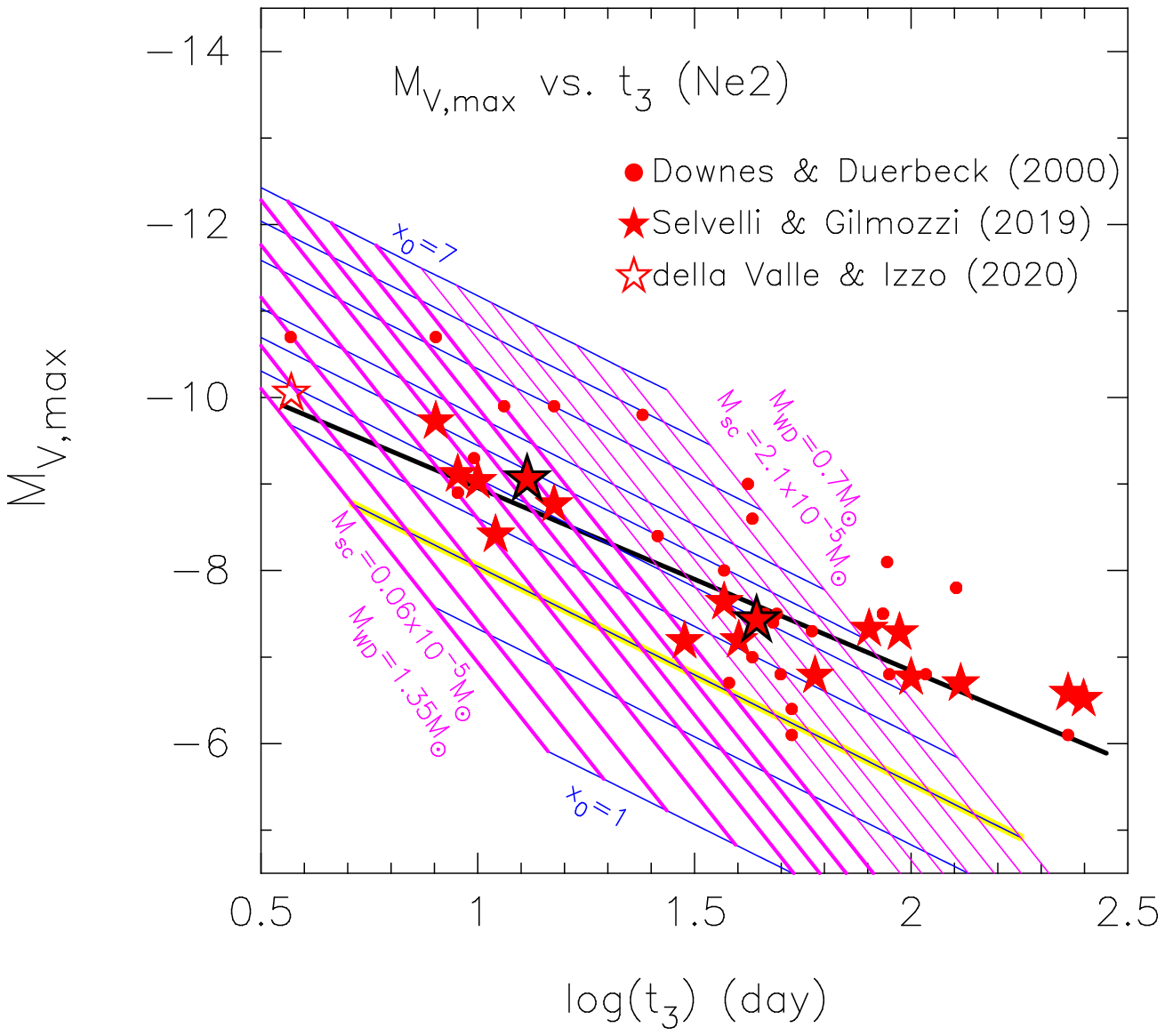}
%\plotone{max_t3_downes_selvelle_crit_x0_ne2.epsi}
%\plotfiddle{evolution1.ps}{5.0cm}{270}{0.4}{0.4}{-170}{220}
\caption{
Same as Figure \ref{max_t3_downes_selvelle_crit_x0}, but for the chemical
composition of Ne nova 2 (Ne2).
Each scaling mass (magenta line) corresponds to the WD mass,
from right to left, $0.7~M_\sun$ to $1.35~M_\sun$ by $0.05~M_\sun$ step
except $1.33~M_\sun$,
the same as those tabulated in Table \ref{critical_mass_ne2_ne3}(Ne2).
The thick solid magenta lines of equi-$M_{\rm sc}$ 
(or equi-$M_{\rm WD}$) correspond to
the WD mass region for neon novae ($M_{\rm WD} \ge 1.05~M_\sun$).
\label{max_t3_downes_selvelle_crit_x0_ne2}}
\end{figure*}

\section{Discussion}
\label{discussion}

%\subsection{Dependence on the chemical composition}
%\label{discussion_chemical_composition}
We have obtained the relation between the peak $V$ brightness 
$M_{V, \rm max}$ and the rate of decline $t_3$ (or $t_2$) in 
Section \ref{peak_brightness_vs_decline} for solar composition
($X=0.70$, $Y=0.28$, and $Z=0.02$ by mass weight).  However, 
it has been reported that nova ejecta are enriched by heavy elements
such as carbon, oxygen, and neon \citep[e.g.,][]{geh98}.  
Here, we discuss the effect of enrichment in heavy elements. 

\citet{kat94h} calculated nova models for various chemical compositions.
We have already tabulated the scaling masses $M_{\rm sc}$
for three cases, CO nova 2 (CO2), CO nova 3 (CO3), and CO nova 4 (CO4) 
in Table \ref{critical_mass_co2_co3_co4} and two cases,
Ne nova 2 (Ne2) and Ne nova 3 (Ne3) in Table \ref{critical_mass_ne2_ne3}.  
The CO3 chemical composition has $X=0.45$, $Y=0.18$, $Z=0.02$, 
$X_{\rm C}=0.15$, and $X_{\rm O}=0.20$ by mass weight.
The Ne2 chemical composition is composed of $X=0.55$, $Y=0.30$, $Z=0.02$,
$X_{\rm O}=0.10$, and $X_{\rm Ne}=0.03$.  We regard the CO3 to be 
a typical chemical composition for CO novae and the Ne2 to be a typical
one for neon novae.  The $M_{\rm sc}$ for the CO3 and Ne2 are
plotted in Figure \ref{mass_cr_wd_mass}(a) together with the solar
composition case in Table \ref{critical_mass_solar}.

If the scaling mass is the same for two different
chemical compositions, their timescales are also the same as inferred 
from Equation (\ref{timescaling_mass_ratio}).  We shift the lines
of CO3 and Ne2 in Figure \ref{mass_cr_wd_mass}(a) toward the right
by $+0.19~M_\sun$ and $+0.12~M_\sun$, respectively, in Figure
\ref{mass_cr_wd_mass}(b).  Then, the three lines (CO3, Ne2, and Solar)
almost overlap with each other at least until $M_{\rm WD}=1.2~M_\sun$.
This means that, for the CO3 case, the timescale ($t_3$) and the peak
$V$ brightness ($M_{V, \rm max}$) of $M_{\rm WD,CO3}+0.19~M_\sun$ 
are the same as those of $M_{\rm WD,solar}$.  In other words,
we have the same MMRD relation between the CO3 and solar compositions at
\begin{equation}
M_{\rm WD,CO3} = M_{\rm WD,solar}-0.19~M_\sun,
\label{wd_mass_relation_co3_solar}
\end{equation}
and between the Ne2 and solar at
\begin{equation}
M_{\rm WD,Ne2} = M_{\rm WD,solar}-0.12~M_\sun.
\label{wd_mass_relation_ne2_solar}
\end{equation}
This difference in the WD mass for different chemical compositions
appears in the WD mass estimate from the light curve fitting
\citep[e.g.,][]{hac16k}.
To confirm these WD mass relations,
we plot the $(\log t_3)$--$M_{V,\rm max}$ diagrams for 
the CO3 and Ne2 cases in Figures
\ref{max_t3_downes_selvelle_crit_x0_co3} and
\ref{max_t3_downes_selvelle_crit_x0_ne2}, respectively,
which are the same as
Figure \ref{max_t3_downes_selvelle_crit_x0} but for different
set of $M_{\rm sc}$.

To better understand these WD mass relations, we examine the case of 
V533~Her (Nova Her 1963).  \citet{sel19} obtained $t_3=44\pm 2$~days and
$M_{V, \rm max}= -7.42\pm 0.22$.  The MMRD point of V533~Her 
is the lower-right outlined red star in Figures
\ref{max_t3_downes_selvelle_crit_x0} and
\ref{max_t3_downes_all_saio_kato2014}.  The position is
on the $M_{\rm WD,solar}= 0.99\pm 0.03~M_\sun$ 
and $t_{\rm rec}= 4000\pm 1000$~yr 
in Figure \ref{max_t3_downes_all_saio_kato2014}.
If we assume the chemical composition to be CO3 for V533~Her,
we obtained the WD mass of 
$M_{\rm WD,CO3}= M_{\rm WD,solar}-0.19~M_\sun = 0.80\pm 0.03~M_\sun$.
For the Ne2 case, we obtain 
$M_{\rm WD,Ne2}= M_{\rm WD,solar}-0.12~M_\sun = 0.87\pm 0.03~M_\sun$.
We can confirm that these two WD masses are reasonable in Figures 
\ref{max_t3_downes_selvelle_crit_x0_co3} and
\ref{max_t3_downes_selvelle_crit_x0_ne2}, respectively.

For the case of $M_{\rm WD,solar}\gtrsim 1.2~M_\sun$, the WD mass 
difference between $M_{\rm WD,solar}$ and $M_{\rm WD,CO3}$ or between
$M_{\rm WD,solar}$ and $M_{\rm WD,Ne2}$ for the same $M_{\rm sc}$ 
becomes smaller as shown in Figure \ref{mass_cr_wd_mass}(a).
Then, the WD mass difference should be measured directly from
Figure \ref{mass_cr_wd_mass}(a).  
We examine the case of GK~Per (Nova Per 1901).  \citet{sel19} obtained
$t_3= 13\pm 1$~days and $M_{V, \rm max}= -9.05\pm 0.16$.
The MMRD point is located at/near $M_{\rm WD,solar}= 1.21~M_\sun$
and $t_{\rm rec}= 8\times 10^4$~yr 
in Figure \ref{max_t3_downes_all_saio_kato2014}. 
If we assume the CO3 chemical composition
for GK~Per, the WD mass difference is estimated to be $0.17~M_\sun$
at $M_{\rm WD,solar}= 1.21~M_\sun$ from Figure \ref{mass_cr_wd_mass}(a).   
Then, we have $M_{\rm WD,CO3}= M_{\rm WD,solar}-0.17~M_\sun = 1.04~M_\sun$.
For the Ne2 case, we obtain the difference of $0.10~M_\sun$ and the
WD mass of $M_{\rm WD,Ne2}= M_{\rm WD,solar}-0.10~M_\sun = 1.11~M_\sun$.
We can also confirm that these two WD masses are reasonable in Figures 
\ref{max_t3_downes_selvelle_crit_x0_co3} and
\ref{max_t3_downes_selvelle_crit_x0_ne2}, respectively.

\citet{hac07k} estimated the WD mass to be 
$M_{\rm WD,Ne2}= 1.15\pm 0.05~M_\sun$ for GK~Per based on the nova 
model light curve fitting.  This is approximately consistent with
the above estimate from the MMRD point.
On the other hand, \citet{hac19ka} estimated the WD mass of V533~Her
to be $M_{\rm WD,Ne2}= 1.03\pm 0.05~M_\sun$ from the nova model light
curve fitting.  This value is larger than the above estimate
from the MMRD relation, $M_{\rm WD,Ne2}= 0.87\pm 0.03~M_\sun$.
This is partly because observational $V$ data are very poor near
the peak and there are jitters with a 0.5 mag amplitude at 2-mag decay
from the peak \citep[see, e.g., Figure 33 of][]{hac19ka}.


%% Included in this acknowledgments section are examples of the
%% AASTeX hypertext markup commands. Use \url without the optional [HREF]
%% argument when you want to print the url directly in the text. Otherwise,
%% use either \url or \anchor, with the HREF as the first argument and the
%% text to be printed in the second.

\section{Conclusions}
\label{conclusions}
\citet{hac06kb, hac10k} constructed free-free emission model light curves
of novae based on the optically-thick wind theory \citep{kat94h}.  
These light curves provided good fit to those of many observed novae.
\citet{hac06kb} also showed that the theoretical nova
light curves are homologous against a normalized time $\tau = t/f_{\rm s}$
and can be expressed with one-parameter 
family of the timescaling factor $f_{\rm s}$.  We find the new
parameter set of $(M_{\rm sc}, x)$ equivalent to $(f_{\rm s}, \tau)$.
Using this equivalent conversion,
we propose a theory for the maximum magnitude versus rate of decline 
(MMRD) relation for novae based on the universal decline law. 
It should be noted that our MMRD relations from the universal decline
law are applicable only to specific light curves of novae such
as S-types defined by \citet{str10}.
Our main conclusions are as follows:\\

\noindent
{\bf 1.} 
We adopt the dimensionless envelope mass $x\equiv M_{\rm env}/M_{\rm sc}$, 
as the parameter that describes the nova light curves. 
The peak $V$ brightness is expressed by an analytic form of $x_0$ and 
$M_{\rm sc}$, i.e., Equation (\ref{approximate_peak_envelope_mass_simple}),
where $x_0$ is the ratio of the initial envelope mass and $M_{\rm sc}$,
$x_0= M_{\rm env, 0}/M_{\rm sc}$.  The decline rate of $t_3$ (or $t_2$) is
also calculated from an analytic formula of $x_0$ and $M_{\rm sc}$, i.e.,
Equation (\ref{t3_time_simple}) (or Equation (\ref{t2_t3_relation})). \\

\noindent
{\bf 2.}
The scaling masses $M_{\rm sc}$ are defined by the envelope mass
having a wind mass-loss rate of
$\log (\dot M_{\rm wind}/M_\sun{\rm ~yr}^{-1}) = -4.7$, which are  
summarized in Tables \ref{critical_mass_co2_co3_co4},
\ref{critical_mass_ne2_ne3}, and \ref{critical_mass_solar}
for various WD masses and chemical compositions
\citep[taken from ][]{kat94h}, and
$x_0$, $M_{V, \rm max}$, $t_2$, and $t_3$ are tabulated in Table
\ref{mmrd_relation_kato2014} for various WD masses 
and mass accretion rates of solar abundance material
\citep[taken from ][with additional calculation for the present 
work]{kat14shn}. \\

\noindent
{\bf 3.} 
We plot our model MMRD relations for the same WD mass, 
same mass accretion rate, and same recurrence time (solar composition).  
The theoretical range of MMRD for expected nova parameters 
well constrains the MMRD points of observed novae 
obtained by \citet{dow00} and \citet{sel19}
except the very long $t_3\gtrsim 200$~days region.
The WD mass ranges mainly from $1.35~M_\sun$ to $0.5~M_\sun$.
The mass accretion rate is typically between 
$3\times 10^{-8}~M_\sun$~yr$^{-1}$ and 
$1\times 10^{-11}~M_\sun$~yr$^{-1}$ centered at
$5\times 10^{-9}~M_\sun$~yr$^{-1}$ in the longer $t_3$ region of
$t_3 > 30$ days. %%%%%%$M_{V, \rm max}\ge -8.0$.  
The recurrence time is typically between 300 yr and $10^6$ yr. \\
%The global trend of an MMRD distribution is consistent with
%the equi-$\dot M_{\rm acc}$ line of
%$5\times 10^{-9}~M_\sun$~yr$^{-1}$. \\

\noindent
{\bf 4.} 
From the MMRD point of an observed nova, we are able to broadly specify 
the WD mass and mass accretion rate.  The main trend in the observed 
MMRD distribution corresponds to the mass accretion
rate of $\dot M_{\rm acc} \sim 5\times 10^{-9}~M_\sun$~yr$^{-1}$ with
the different WD masses in the longer $t_3$ region of $t_3 > 30$ days.  
The scatter of nova MMRD points from the main trend line
can be attributed to a large scatter of the observed
mass accretion rates from that of the main trend defined by 
$\dot M_{\rm acc} \sim 5\times 10^{-9}~M_\sun$~yr$^{-1}$.
{\it
Thus, the global trend of an MMRD relation does exist,
but its scatter is too large for it to be a precision distance indicator
of individual novae.
}
\\

\noindent
{\bf 5.} 
In the shorter $t_3$ region of $t_3 < 30$~days,
%%%$M_{V, \rm max}\le -8.0$,
the main trend MMRD relation traverses the four lines of 
mass-accretion rates, 
$\dot M_{\rm acc}= 3\times 10^{-9}~M_\sun$~yr$^{-1}$,
$10^{-9}~M_\sun$~yr$^{-1}$,
$10^{-10}~M_\sun$~yr$^{-1}$, and
$10^{-11}~M_\sun$~yr$^{-1}$.
In general, the smaller the $t_3$ time, the smaller the $\dot M_{\rm acc}$.  
The recurrence time is typically between $t_{\rm rec}=10000$ yr and $10^6$ yr.
%The global trend of an MMRD distribution is consistent with
%the existing novae.
\\

\noindent
{\bf 6.} 
If we divide the data of \citet{sel19} into two groups
by the brightness of $M_{V, \rm max} = -8.0$, the upper brighter group
broadly correspond to cataclysmic variables below the period gap 
($\dot M_{\rm acc}\sim 10^{-10}~M_\sun$~yr$^{-1}$) while
the lower fainter group correspond to binaries above the period gap
($\dot M_{\rm acc}\sim 5 \times 10^{-9}~M_\sun$~yr$^{-1}$).
\\

\noindent
{\bf 7.} 
The lower the mass accretion rate, the larger the ignition mass for
a given WD mass and chemical composition.  Thus, the lower the mass
accretion rate, the brighter the $V$ peak of a nova.
There is a $M_V \sim -10.4$ mag cap for the maximum brightness 
if we limit the mass accretion rate above $\dot M_{\rm acc} 
\ge 1\times 10^{-11} ~M_\sun$~yr$^{-1}$ \citep[e.g.,][]{kni11}.  
This explains that the brightest peaks of novae are at/around 
$M_V \sim -10.4$.
{\it Thus, we clarified the reason for the brightness cap
($M_V \sim -10.4$) of classical novae, which is constrained
by the lowest mass-accretion rate of $\dot M_{\rm acc} \sim 
1\times 10^{-11} ~M_\sun$~{\rm yr}$^{-1}$. 
}
\\

\noindent
{\bf 8.} 
Finally, we discussed the effect of enrichment of heavy elements
in nova ejecta.  The enrichment has the same effect as the WD mass decrease.
For example, the difference between the abundances of solar and Ne2 novae 
is 0.12 $M_\sun$, i.e., $M_{\rm WD, solar}= M_{\rm WD, Ne2}
+ 0.12~M_\sun$, for the same $M_{V,\rm max}$ and $t_3$.
\\

\acknowledgments
     I.H. and M.K. thank 
Department of Astronomy, San Diego State University,
for the warm hospitality, 
during which we initiated the present work.
We are grateful to the anonymous referee for useful comments,
which improved the manuscript.



\appendix

\section{Nova Ignition Models}
\label{accreted_envelope_ignition}

We have calculated the MMRD points, $M_{V,\rm max}$ and $t_3$ (or $t_2$),
based on the result of \citet{kat14shn} for various WD masses and 
mass accretion rates (solar abundance) together with additional calculation
that expanded the parameter range of mass-accretion rate   
for the present work.  Corresponding parameters are the WD mass, 
$M_{\rm WD}$, mass-accretion rate, $\dot M_{\rm acc}$, 
accumulation mass, $M_{\rm acc}$ (Section \ref{ignition_mass_model}),
scaling mass, $M_{\rm sc}$ (Section \ref{timescale_light_curves}),
parameter $x_0$ (Sections \ref{timescale_light_curves} and
\ref{universal_paek_brightness}),
peak absolute $V$ magnitude, $M_{V, \rm max}$ (Equation
(\ref{approximate_peak_envelope_mass_simple})), 
times $t_2$, $t_3$ (Section \ref{approximate_relation}, 
Equation (\ref{t3_time_simple})), 
and the recurrence time, $t_{\rm rec}$ 
(Section \ref{peak_brightness_vs_decline}, 
Figure \ref{max_t3_downes_all_saio_kato2014}).

%Table 4 

\startlongtable
\begin{deluxetable*}{lllllllll}
\tabletypesize{\scriptsize}
%%%\rotate
\tablecaption{MMRD Relation for Kato et al.'s Model\tablenotemark{a}
\label{mmrd_relation_kato2014}}
\tablewidth{0pt}
\tablehead{
\colhead{$M_{\rm WD}$} & \colhead{$\dot M_{\rm acc}$} 
& \colhead{$M_{\rm acc}$} & \colhead{$M_{\rm sc}$} 
& \colhead{$x_0$} & \colhead{$M_{V, \rm max}$} 
& \colhead{$t_2$} & \colhead{$t_3$} & \colhead{$t_{\rm rec}$} \\
\colhead{($M_\sun$)} & \colhead{($M_\sun$~yr$^{-1}$)} & \colhead{($M_\sun$)} 
& \colhead{($M_\sun$)} & \colhead{} & \colhead{} 
& \colhead{(days)} & \colhead{(days)} & \colhead{(yr)} }
\startdata
0.6 & 1.0E-11 & 2.52E-4 & 5.88E-5 & 4.56 & -7.20 & 78.6 & 146. & 2.52E+7 \\
0.6 & 3.0E-11 & 2.42E-4 & 5.88E-5 & 4.39 & -7.04 & 83.2 & 154. & 8.08E+6 \\
0.6 & 5.0E-11 & 2.39E-4 & 5.88E-5 & 4.33 & -6.99 & 84.9 & 157. & 4.78E+6 \\
0.6 & 1.0E-10 & 2.37E-4 & 5.88E-5 & 4.31 & -6.97 & 85.6 & 159. & 2.37E+6 \\
0.6 & 3.0E-10 & 2.31E-4 & 5.88E-5 & 4.21 & -6.87 & 88.8 & 164. & 7.71E+5 \\
0.6 & 1.0E-9 & 2.17E-4 & 5.88E-5 & 3.97 & -6.62 & 97.0 & 180. & 2.17E+5 \\
0.6 & 1.6E-9 & 2.07E-4 & 5.88E-5 & 3.79 & -6.44 & 104. & 192. & 1.29E+5 \\
0.6 & 3.0E-9 & 1.91E-4 & 5.88E-5 & 3.52 & -6.12 & 116. & 215. & 6.35E+4 \\
0.6 & 5.0E-9 & 1.75E-4 & 5.88E-5 & 3.24 & -5.79 & 131. & 243. & 3.49E+4 \\
0.6 & 1.0E-8 & 1.48E-4 & 5.88E-5 & 2.78 & -5.16 & 165. & 306. & 1.48E+4 \\
0.6 & 1.6E-8 & 1.32E-4 & 5.88E-5 & 2.52 & -4.74 & 192. & 356. & 8235. \\
0.6 & 2.0E-8 & 1.25E-4 & 5.88E-5 & 2.41 & -4.56 & 205. & 380. & 6269. \\
0.6 & 3.0E-8 & 1.15E-4 & 5.88E-5 & 2.24 & -4.26 & 229. & 424. & 3846. \\
\\
0.7 & 1.0E-11 & 1.72E-4 & 3.35E-5 & 5.44 & -8.54 & 34.4 & 63.6 & 1.72E+7 \\
0.7 & 3.0E-11 & 1.63E-4 & 3.35E-5 & 5.17 & -8.32 & 37.1 & 68.8 & 5.44E+6 \\
0.7 & 5.0E-11 & 1.60E-4 & 3.35E-5 & 5.06 & -8.24 & 38.3 & 70.9 & 3.19E+6 \\
0.7 & 1.0E-10 & 1.55E-4 & 3.35E-5 & 4.93 & -8.13 & 39.8 & 73.7 & 1.55E+6 \\
0.7 & 3.0E-10 & 1.48E-4 & 3.35E-5 & 4.70 & -7.93 & 42.8 & 79.2 & 4.92E+5 \\
0.7 & 1.0E-9 & 1.35E-4 & 3.35E-5 & 4.33 & -7.59 & 48.4 & 89.6 & 1.35E+5 \\
0.7 & 3.0E-9 & 1.10E-4 & 3.35E-5 & 3.59 & -6.82 & 64.1 & 119. & 3.68E+4 \\
0.7 & 1.0E-8 & 8.81E-5 & 3.35E-5 & 2.92 & -5.98 & 87.2 & 161. & 8808. \\
0.7 & 3.0E-8 & 6.47E-5 & 3.35E-5 & 2.22 & -4.85 & 131. & 243. & 2156. \\
0.7 & 5.0E-8 & 5.68E-5 & 3.35E-5 & 1.99 & -4.39 & 155. & 288. & 1136. \\
0.7 & 6.0E-8 & 5.47E-5 & 3.35E-5 & 1.93 & -4.25 & 163. & 302. & 911. \\
\\
0.8 & 1.0E-11 & 1.18E-4 & 2.22E-5 & 5.56 & -9.07 & 22.1 & 40.9 & 1.18E+7 \\
0.8 & 3.0E-11 & 1.10E-4 & 2.22E-5 & 5.20 & -8.80 & 24.4 & 45.2 & 3.68E+6 \\
0.8 & 5.0E-11 & 1.07E-4 & 2.22E-5 & 5.06 & -8.68 & 25.4 & 47.1 & 2.15E+6 \\
0.8 & 1.0E-10 & 1.06E-4 & 2.22E-5 & 4.98 & -8.62 & 26.1 & 48.3 & 1.06E+6 \\
0.8 & 3.0E-10 & 1.00E-4 & 2.22E-5 & 4.74 & -8.41 & 28.1 & 52.0 & 3.34E+5 \\
0.8 & 1.0E-9 & 9.18E-5 & 2.22E-5 & 4.37 & -8.07 & 31.8 & 58.8 & 9.18E+4 \\
0.8 & 1.6E-9 & 8.70E-5 & 2.22E-5 & 4.15 & -7.86 & 34.3 & 63.5 & 5.44E+4 \\
0.8 & 3.0E-9 & 7.90E-5 & 2.22E-5 & 3.79 & -7.49 & 39.3 & 72.8 & 2.63E+4 \\
0.8 & 5.0E-9 & 7.21E-5 & 2.22E-5 & 3.48 & -7.14 & 44.7 & 82.7 & 1.44E+4 \\
0.8 & 1.0E-8 & 6.01E-5 & 2.22E-5 & 2.94 & -6.44 & 57.6 & 107. & 6006. \\
0.8 & 1.6E-8 & 5.19E-5 & 2.22E-5 & 2.57 & -5.89 & 70.3 & 130. & 3244. \\
0.8 & 3.0E-8 & 4.29E-5 & 2.22E-5 & 2.17 & -5.18 & 90.9 & 168. & 1430. \\
0.8 & 5.0E-8 & 3.67E-5 & 2.22E-5 & 1.89 & -4.61 & 112. & 207. & 734. \\
0.8 & 7.0E-8 & 3.34E-5 & 2.22E-5 & 1.74 & -4.27 & 127. & 234. & 477. \\
0.8 & 7.5E-8 & 3.28E-5 & 2.22E-5 & 1.71 & -4.21 & 130. & 240. & 437. \\
\\
0.9 & 1.0E-11 & 7.93E-5 & 1.39E-5 & 5.90 & -9.83 & 12.7 & 23.4 & 7.93E+6 \\
0.9 & 3.0E-11 & 7.29E-5 & 1.39E-5 & 5.44 & -9.49 & 14.3 & 26.5 & 2.43E+6 \\
0.9 & 5.0E-11 & 7.04E-5 & 1.39E-5 & 5.26 & -9.35 & 15.0 & 27.8 & 1.41E+6 \\
0.9 & 1.0E-10 & 6.74E-5 & 1.39E-5 & 5.05 & -9.18 & 16.0 & 29.6 & 6.74E+5 \\
0.9 & 3.0E-10 & 6.31E-5 & 1.39E-5 & 4.74 & -8.92 & 17.6 & 32.6 & 2.10E+5 \\
0.9 & 1.0E-9 & 5.66E-5 & 1.39E-5 & 4.27 & -8.49 & 20.5 & 38.0 & 5.66E+4 \\
0.9 & 3.0E-9 & 4.84E-5 & 1.39E-5 & 3.69 & -7.88 & 25.6 & 47.5 & 1.61E+4 \\
0.9 & 1.0E-8 & 3.87E-5 & 1.39E-5 & 2.99 & -7.01 & 35.2 & 65.1 & 3867. \\
0.9 & 3.0E-8 & 2.72E-5 & 1.39E-5 & 2.17 & -5.69 & 56.9 & 105. & 908. \\
0.9 & 5.0E-8 & 2.30E-5 & 1.39E-5 & 1.86 & -5.06 & 71.6 & 133. & 459. \\
0.9 & 7.0E-8 & 2.06E-5 & 1.39E-5 & 1.69 & -4.66 & 82.7 & 153. & 294. \\
0.9 & 9.0E-8 & 1.90E-5 & 1.39E-5 & 1.58 & -4.38 & 91.6 & 170. & 212. \\
0.9 & 1.0E-7 & 1.84E-5 & 1.39E-5 & 1.53 & -4.27 & 95.5 & 177. & 184. \\
0.9 & 1.1E-7 & 1.79E-5 & 1.39E-5 & 1.50 & -4.17 & 99.0 & 183. & 163. \\
\\
1.0 & 1.0E-11 & 5.07E-5 & 9.25E-6 & 5.68 & -10.11 & 8.89 & 16.5 & 5.07E+6 \\
1.0 & 3.0E-11 & 4.59E-5 & 9.25E-6 & 5.16 & -9.71 & 10.3 & 19.1 & 1.53E+6 \\
1.0 & 5.0E-11 & 4.41E-5 & 9.25E-6 & 4.96 & -9.55 & 10.9 & 20.2 & 8.81E+5 \\
1.0 & 1.0E-10 & 4.30E-5 & 9.25E-6 & 4.85 & -9.46 & 11.3 & 20.9 & 4.30E+5 \\
1.0 & 3.0E-10 & 4.00E-5 & 9.25E-6 & 4.53 & -9.18 & 12.5 & 23.2 & 1.33E+5 \\
1.0 & 1.0E-9 & 3.63E-5 & 9.25E-6 & 4.12 & -8.79 & 14.4 & 26.6 & 3.63E+4 \\
1.0 & 1.6E-9 & 3.44E-5 & 9.25E-6 & 3.92 & -8.58 & 15.5 & 28.7 & 2.15E+4 \\
1.0 & 3.0E-9 & 3.12E-5 & 9.25E-6 & 3.57 & -8.20 & 17.9 & 33.1 & 1.04E+4 \\
1.0 & 5.0E-9 & 2.85E-5 & 9.25E-6 & 3.27 & -7.84 & 20.3 & 37.7 & 5689. \\
1.0 & 1.0E-8 & 2.39E-5 & 9.25E-6 & 2.78 & -7.16 & 26.0 & 48.2 & 2386. \\
1.0 & 1.6E-8 & 2.05E-5 & 9.25E-6 & 2.41 & -6.58 & 32.2 & 59.6 & 1280. \\
1.0 & 3.0E-8 & 1.67E-5 & 9.25E-6 & 2.00 & -5.80 & 42.6 & 78.9 & 555. \\
1.0 & 5.0E-8 & 1.39E-5 & 9.25E-6 & 1.69 & -5.12 & 54.6 & 101. & 277. \\
1.0 & 1.0E-7 & 1.09E-5 & 9.25E-6 & 1.37 & -4.25 & 75.0 & 139. & 109. \\
1.0 & 1.2E-7 & 1.02E-5 & 9.25E-6 & 1.30 & -4.03 & 81.1 & 150. & 85.1 \\
\\
1.1 & 1.0E-11 & 3.01E-5 & 5.81E-6 & 5.36 & -10.37 & 6.10 & 11.3 & 3.01E+6 \\
1.1 & 3.0E-11 & 2.68E-5 & 5.81E-6 & 4.78 & -9.91 & 7.23 & 13.4 & 8.93E+5 \\
1.1 & 5.0E-11 & 2.56E-5 & 5.81E-6 & 4.58 & -9.73 & 7.72 & 14.3 & 5.12E+5 \\
1.1 & 1.0E-10 & 2.49E-5 & 5.81E-6 & 4.46 & -9.62 & 8.03 & 14.9 & 2.49E+5 \\
1.1 & 3.0E-10 & 2.31E-5 & 5.81E-6 & 4.14 & -9.31 & 8.98 & 16.6 & 7.69E+4 \\
1.1 & 1.0E-9 & 2.09E-5 & 5.81E-6 & 3.76 & -8.92 & 10.4 & 19.2 & 2.09E+4 \\
1.1 & 1.6E-9 & 1.98E-5 & 5.81E-6 & 3.57 & -8.70 & 11.2 & 20.8 & 1.23E+4 \\
1.1 & 3.0E-9 & 1.79E-5 & 5.81E-6 & 3.26 & -8.32 & 12.9 & 23.8 & 5976. \\
1.1 & 5.0E-9 & 1.64E-5 & 5.81E-6 & 2.99 & -7.97 & 14.6 & 27.1 & 3277. \\
1.1 & 1.0E-8 & 1.38E-5 & 5.81E-6 & 2.55 & -7.31 & 18.6 & 34.5 & 1381. \\
1.1 & 1.6E-8 & 1.19E-5 & 5.81E-6 & 2.22 & -6.73 & 23.0 & 42.5 & 742. \\
1.1 & 3.0E-8 & 9.62E-6 & 5.81E-6 & 1.83 & -5.94 & 30.6 & 56.7 & 321. \\
1.1 & 5.0E-8 & 7.95E-6 & 5.81E-6 & 1.54 & -5.23 & 39.6 & 73.4 & 159. \\
1.1 & 1.0E-7 & 6.13E-6 & 5.81E-6 & 1.23 & -4.29 & 55.8 & 103. & 61.3 \\
1.1 & 1.6E-7 & 5.14E-6 & 5.81E-6 & 1.06 & -3.68 & 69.7 & 129. & 32.2 \\
\\
1.2 & 1.0E-11 & 1.54E-5 & 3.60E-6 & 4.42 & -10.10 & 5.04 & 9.34 & 1.54E+6 \\
1.2 & 3.0E-11 & 1.35E-5 & 3.60E-6 & 3.90 & -9.59 & 6.08 & 11.3 & 4.50E+5 \\
1.2 & 5.0E-11 & 1.28E-5 & 3.60E-6 & 3.70 & -9.37 & 6.59 & 12.2 & 2.55E+5 \\
1.2 & 1.0E-10 & 1.24E-5 & 3.60E-6 & 3.61 & -9.26 & 6.85 & 12.7 & 1.24E+5 \\
1.2 & 3.0E-10 & 1.14E-5 & 3.60E-6 & 3.33 & -8.93 & 7.73 & 14.3 & 3.81E+4 \\
1.2 & 1.0E-9 & 1.03E-5 & 3.60E-6 & 3.02 & -8.53 & 8.94 & 16.6 & 1.03E+4 \\
1.2 & 1.6E-9 & 9.77E-6 & 3.60E-6 & 2.87 & -8.31 & 9.67 & 17.9 & 6107. \\
1.2 & 3.0E-9 & 8.89E-6 & 3.60E-6 & 2.62 & -7.94 & 11.1 & 20.5 & 2962. \\
1.2 & 5.0E-9 & 8.12E-6 & 3.60E-6 & 2.41 & -7.59 & 12.6 & 23.4 & 1624. \\
1.2 & 1.0E-8 & 6.85E-6 & 3.60E-6 & 2.05 & -6.94 & 15.9 & 29.5 & 685. \\
1.2 & 1.6E-8 & 5.89E-6 & 3.60E-6 & 1.79 & -6.37 & 19.6 & 36.3 & 368. \\
1.2 & 3.0E-8 & 4.76E-6 & 3.60E-6 & 1.47 & -5.57 & 26.2 & 48.6 & 159. \\
1.2 & 5.0E-8 & 3.90E-6 & 3.60E-6 & 1.24 & -4.85 & 34.1 & 63.2 & 78.0 \\
1.2 & 1.0E-7 & 2.97E-6 & 3.60E-6 & 0.978 & -3.88 & 48.5 & 89.9 & 29.7 \\
1.2 & 1.6E-7 & 2.45E-6 & 3.60E-6 & 0.832 & -3.21 & 61.8 & 114. & 15.3 \\
1.2 & 1.8E-7 & 2.33E-6 & 3.60E-6 & 0.800 & -3.05 & 65.6 & 122. & 12.9 \\
\\
1.25 & 1.0E-11 & 9.93E-6 & 2.61E-6 & 3.95 & -9.99 & 4.33 & 8.01 & 9.93E+5 \\
1.25 & 3.0E-11 & 8.61E-6 & 2.61E-6 & 3.44 & -9.42 & 5.32 & 9.84 & 2.87E+5 \\
1.25 & 5.0E-11 & 8.15E-6 & 2.61E-6 & 3.27 & -9.20 & 5.75 & 10.7 & 1.63E+5 \\
1.25 & 1.0E-10 & 7.92E-6 & 2.61E-6 & 3.18 & -9.09 & 5.99 & 11.1 & 7.92E+4 \\
1.25 & 3.0E-10 & 7.27E-6 & 2.61E-6 & 2.93 & -8.75 & 6.78 & 12.6 & 2.42E+4 \\
1.25 & 1.0E-9 & 6.57E-6 & 2.61E-6 & 2.66 & -8.35 & 7.84 & 14.5 & 6565. \\
1.25 & 1.6E-9 & 6.21E-6 & 2.61E-6 & 2.52 & -8.13 & 8.48 & 15.7 & 3881. \\
1.25 & 3.0E-9 & 5.65E-6 & 2.61E-6 & 2.31 & -7.77 & 9.69 & 17.9 & 1884. \\
1.25 & 5.0E-9 & 5.16E-6 & 2.61E-6 & 2.12 & -7.42 & 11.0 & 20.4 & 1031. \\
1.25 & 1.0E-8 & 4.34E-6 & 2.61E-6 & 1.81 & -6.76 & 14.0 & 25.9 & 434. \\
1.25 & 1.6E-8 & 3.75E-6 & 2.61E-6 & 1.58 & -6.20 & 17.1 & 31.7 & 234. \\
1.25 & 3.0E-8 & 3.02E-6 & 2.61E-6 & 1.30 & -5.40 & 22.9 & 42.4 & 101. \\
1.25 & 5.0E-8 & 2.49E-6 & 2.61E-6 & 1.10 & -4.70 & 29.6 & 54.8 & 49.8 \\
1.25 & 1.0E-7 & 1.87E-6 & 2.61E-6 & 0.858 & -3.69 & 42.8 & 79.2 & 18.7 \\
1.25 & 1.6E-7 & 1.52E-6 & 2.61E-6 & 0.722 & -2.98 & 55.3 & 103. & 9.47 \\
1.25 & 2.0E-7 & 1.37E-6 & 2.61E-6 & 0.668 & -2.65 & 62.3 & 115. & 6.86 \\
\\
1.3 & 1.0E-11 & 5.59E-6 & 1.74E-6 & 3.35 & -9.75 & 3.69 & 6.84 & 5.59E+5 \\
1.3 & 3.0E-11 & 4.83E-6 & 1.74E-6 & 2.91 & -9.17 & 4.56 & 8.45 & 1.61E+5 \\
1.3 & 5.0E-11 & 4.55E-6 & 1.74E-6 & 2.75 & -8.94 & 4.96 & 9.18 & 9.11E+4 \\
1.3 & 1.0E-10 & 4.40E-6 & 1.74E-6 & 2.66 & -8.80 & 5.21 & 9.66 & 4.40E+4 \\
1.3 & 3.0E-10 & 4.03E-6 & 1.74E-6 & 2.45 & -8.46 & 5.90 & 10.9 & 1.34E+4 \\
1.3 & 1.0E-9 & 3.65E-6 & 1.74E-6 & 2.23 & -8.07 & 6.80 & 12.6 & 3646. \\
1.3 & 1.5E-9 & 3.48E-6 & 1.74E-6 & 2.13 & -7.89 & 7.26 & 13.5 & 2319. \\
1.3 & 3.0E-9 & 3.14E-6 & 1.74E-6 & 1.94 & -7.49 & 8.39 & 15.5 & 1046. \\
1.3 & 1.0E-8 & 2.41E-6 & 1.74E-6 & 1.52 & -6.49 & 12.1 & 22.3 & 241. \\
1.3 & 3.0E-8 & 1.69E-6 & 1.74E-6 & 1.10 & -5.16 & 19.6 & 36.2 & 56.2 \\
1.3 & 1.0E-7 & 1.02E-6 & 1.74E-6 & 0.720 & -3.40 & 37.1 & 68.7 & 10.2 \\
1.3 & 1.6E-7 & 8.27E-7 & 1.74E-6 & 0.609 & -2.71 & 47.7 & 88.3 & 5.17 \\
1.3 & 2.0E-7 & 7.44E-7 & 1.74E-6 & 0.561 & -2.38 & 53.9 & 99.8 & 3.72 \\
1.3 & 2.2E-7 & 7.12E-7 & 1.74E-6 & 0.543 & -2.24 & 56.6 & 105. & 3.24 \\
\\
1.35 & 1.0E-11 & 2.38E-6 & 9.40E-7 & 2.64 & -9.43 & 2.86 & 5.30 & 2.38E+5 \\
1.35 & 3.0E-11 & 2.06E-6 & 9.40E-7 & 2.30 & -8.86 & 3.51 & 6.50 & 6.86E+4 \\
1.35 & 5.0E-11 & 1.95E-6 & 9.40E-7 & 2.19 & -8.66 & 3.79 & 7.01 & 3.91E+4 \\
1.35 & 1.0E-10 & 1.85E-6 & 9.40E-7 & 2.08 & -8.45 & 4.09 & 7.57 & 1.85E+4 \\
1.35 & 3.0E-10 & 1.71E-6 & 9.40E-7 & 1.93 & -8.14 & 4.57 & 8.47 & 5700. \\
1.35 & 1.0E-9 & 1.54E-6 & 9.40E-7 & 1.75 & -7.74 & 5.28 & 9.78 & 1544. \\
1.35 & 1.6E-9 & 1.46E-6 & 9.40E-7 & 1.66 & -7.53 & 5.70 & 10.6 & 914. \\
1.35 & 3.0E-9 & 1.34E-6 & 9.40E-7 & 1.53 & -7.18 & 6.47 & 12.0 & 445. \\
1.35 & 5.0E-9 & 1.22E-6 & 9.40E-7 & 1.41 & -6.84 & 7.33 & 13.6 & 244. \\
1.35 & 1.0E-8 & 1.03E-6 & 9.40E-7 & 1.21 & -6.21 & 9.21 & 17.1 & 103. \\
1.35 & 1.6E-8 & 9.06E-7 & 9.40E-7 & 1.07 & -5.72 & 11.0 & 20.4 & 56.6 \\
1.35 & 3.0E-8 & 7.33E-7 & 9.40E-7 & 0.889 & -4.94 & 14.6 & 27.1 & 24.4 \\
1.35 & 5.0E-8 & 6.03E-7 & 9.40E-7 & 0.751 & -4.24 & 18.8 & 34.9 & 12.1 \\
1.35 & 1.0E-7 & 4.50E-7 & 9.40E-7 & 0.587 & -3.23 & 27.2 & 50.4 & 4.50 \\
1.35 & 1.6E-7 & 3.62E-7 & 9.40E-7 & 0.494 & -2.52 & 35.3 & 65.3 & 2.26 \\
1.35 & 2.0E-7 & 3.25E-7 & 9.40E-7 & 0.455 & -2.18 & 39.9 & 73.9 & 1.62 \\
1.35 & 2.5E-7 & 2.91E-7 & 9.40E-7 & 0.419 & -1.84 & 45.1 & 83.6 & 1.17
\enddata
\tablenotetext{a}{chemical composition of the envelope is assumed
to be that of ``Solar'' in Table 2 of \citet{hac06kb}.}
\end{deluxetable*}

\clearpage


%% If you are not including electronic art with your submission, you may
%% mark up your captions using the \figcaption command. See the 
%% No more than seven \figcaption commands are allowed per page, 
%% so if you have more than seven captions, insert a \clearpage 

%% Tables should be submitted one per page, so put a \clearpage before

%% Two options are available to the author for producing tables:  the
%% deluxetable environment provided by the AASTeX package or the LaTeX
%% table environment.  Use of deluxetable is preferred.
%%

%% Three table samples follow, two marked up in the deluxetable environment,
%% one marked up as a LaTeX table.

%% The following command ends your manuscript. LaTeX will ignore any text

\end{document}